\documentclass[11pt,a4paper]{article}
\pdfoutput=1

\usepackage{jheppub}
\usepackage{latexsym}
\usepackage{multirow}
\usepackage{color}
\usepackage[usenames,dvipsnames,table]{xcolor}
\usepackage{graphicx}
\usepackage{epsfig}  
\usepackage{dcolumn}
\usepackage{bm}
\usepackage{dcolumn}
\usepackage{textcomp}
\usepackage{float}
\usepackage{hypcap}
\usepackage[]{hyperref}
\usepackage{makecell}
\usepackage{multirow}
\usepackage{color}
\usepackage{pifont}

 \usepackage{caption}
  \newcommand{\capdef}{}
  \newcommand{\mycaption}[2][\capdef]{\renewcommand{\capdef}{#2}
       \caption[#1]{{\footnotesize #2}}}
  \newcommand{\be}{\begin{equation}}
   \newcommand{\ee}{\end{equation}}

\hypersetup{
  bookmarks=true,         
  unicode=false,          
  pdftoolbar=true,        
 pdfmenubar=true,        
 pdffitwindow=true,     
 pdfstartview={FitH},    
 pdfsubject={Neutrino Oscillations Phenomenology},   
 pdfnewwindow=true,      
 pdfcreator={RevTeX},
 colorlinks=true,       
 linkcolor=red,          
 citecolor=blue,        
 filecolor=black,      
 urlcolor=blue,           
  }
\preprint{IP/BBSR/2019-03}
\title{Enhancing Sensitivity to Non-Standard Neutrino Interactions at INO 
combining muon and hadron information }
\author[a,b]{Amina Khatun,}
\author[a,f]{Sabya Sachi Chatterjee,} 
\author[c,d]{Tarak Thakore,}
\author[a,b,e]{Sanjib Kumar Agarwalla}

\affiliation[a]{Institute of Physics, Sachivalaya Marg, Sainik School Post, 
Bhubaneswar 751005, India}
\affiliation[b]{Homi Bhabha National Institute, Anushakti Nagar, Mumbai 400085, India}
\affiliation[c]{Louisiana State University,
Baton Rouge, Louisiana, 70803 U.S.A.}
\affiliation[d]{Instituto de F\`{i}sica Corpuscular, CSIC - Universitat de Val\`{e}ncia , c/ Catedr\`{a}tico Jos\`{e} Beltr\`{a}n 2, E-46980 Paterna, Valencia, Spain}
\affiliation[e]{International Centre for Theoretical Physics, Strada Costiera 11, 34151 Trieste, Italy}
\affiliation[f]{Institute for Particle Physics Phenomenology (IPPP), Department of Physics, Durham University, Durham, DH1 3LE, UK}

\emailAdd{amina@iopb.res.in}
\emailAdd{sabya.s.chatterjee@durham.ac.uk}
\emailAdd{tarak.thakore@ific.uv.es}
\emailAdd{sanjib@iopb.res.in}

\abstract{The neutral current non-standard interactions (NSI's) of
neutrino with matter fermions while propagating through long distances 
inside the Earth matter can give rise to the extra matter potentials apart 
from the standard MSW potential due to the $W$-mediated interactions 
in matter. In this paper, we explore the impact of flavor violating  neutral 
current NSI parameter $\varepsilon_{\mu\tau}$ in the oscillation of 
atmospheric neutrino and antineutrino using the 50 kt magnetized ICAL 
detector at INO. We find that due to non-zero $\varepsilon_{\mu\tau}$, 
$\nu_\mu\rightarrow\nu_\mu$ and $\bar\nu_\mu\rightarrow\bar\nu_\mu$ 
transition probabilities get modified substantially at higher energies and 
longer baselines, where vacuum oscillation dominates. We estimate the 
sensitivity of the ICAL detector for various choices of binning schemes 
and observables. The most optimistic bound on $\varepsilon_{\mu\tau}$ 
that we obtain is $-0.01 <  \varepsilon_{\mu\tau} < 0.01 $ at 90$\%$ C.L. 
using 500 kt$\cdot$yr exposure and considering 
$E_\mu,\, \cos\theta_\mu,\,E'_{\rm had}$ as observables in their ranges 
[1,\,21]\,GeV, [-1,\,1], and [0,\,25]\,GeV respectively. For the first time we 
show that the charge identification capability of the ICAL detector is crucial 
to set stringent constraints on $\varepsilon_{\mu\tau}$. We also show that 
when we marginalize over $\varepsilon_{\mu\tau}$ in fit in its range of -0.1 
to 0.1, the mass hierarchy sensitivity deteriorates by 10$\%$ to 20$\%$ 
depending on the analysis mode, and the precision measurements of 
atmospheric parameters remain quite robust at the ICAL detector.
}

\arxivnumber{1907.aaaaa}

\begin{document}
\maketitle
\flushbottom

\section{Introduction }
\label{Introduction}
The observation of neutrino oscillation proves that  the neutrinos are massive and mix with each other, which is the first direct evidence for physics beyond the Standard Model\,\cite{Olive:2016xmw}. Currently, the three-flavor lepton mixing is confirmed and neutrino oscillation has entered the era of precision\,\cite{Capozzi:2017ipn,Esteban:2018azc,deSalas:2017kay}.  
To explain the tiny masses of neutrino and large leptonic mixing angles, the extensions of the Standard Model allow some interactions which are not possible in the Standard Model. These interactions are termed as non-standard interactions (NSI's). 
The presence of NSI's in nature can have subdominant effect on the oscillation of neutrino and antineutrino. Therefore, the phenomenological consequences of NSI's in three-flavor mixing using neutrino oscillation experiments are interesting and widely studied by many authors in Refs.\,\cite{GonzalezGarcia:1998hj,Bergmann:1998rg,Bergmann:1999rz,Lipari:1999vh,Guzzo:2000kx,GonzalezGarcia:2001mp,Gago:2001xg,Berezhiani:2001rs,Huber:2001zw,Fornengo:2001pm,Huber:2002bi,Gago:2001si,Berezhiani:2001rt,Ota:2001pw,Davidson:2003ha,Friedland:2004ah,GonzalezGarcia:2004wg,Guzzo:2004ue,Friedland:2005vy,Kitazawa:2006iq,Mangano:2006ar,Blennow:2005qj,EstebanPretel:2007yu,Barranco:2007tz,Ribeiro:2007ud,Kopp:2007mi,Blennow:2008ym,EstebanPretel:2008qi,Barranco:2007ej,Kopp:2007ne,Blennow:2007pu,Ohlsson:2008gx,Gavela:2008ra,Bolanos:2008km,Antusch:2008tz,Biggio:2009nt,Escrihuela:2009up,Gago:2009ij,Biggio:2009kv,Oki:2010uc,Kopp:2010qt,Forero:2011zz,Escrihuela:2011cf,GonzalezGarcia:2011my,Coloma:2011rq,Adhikari:2012vc,Agarwalla:2012wf,Gonzalez-Garcia:2013usa,Ohlsson:2013epa,Ohlsson:2012kf,Choubey:2015xha,Farzan:2017xzy}.    

In this paper, we study the impact of neutral current (NC) non-standard interactions of neutrino which 
may arise when atmospheric neutrinos travel long distances inside the Earth. While NC NSI's 
affect neutrinos during their propagation, there are charged-current NSI's which may modify 
the neutrino fluxes at the production stage and interaction cross-section at the detection level. 
In this work, we only focus on the NC NSI's, and do not consider NSI's at production and detection 
level. In most of the cases, NSI's come into the picture as a low-energy manifestation of high-energy 
theory involving new heavy states. For a detailed discussion on this topic, see the 
reviews\,\cite{Biggio:2009nt,Ohlsson:2012kf,Miranda:2015dra,Farzan:2017xzy}. 
Therefore, at low energies, a neutral current NSI can be described by a four-fermion dimension-six 
operator\,\cite{Wolfenstein:1977ue}, 
\begin{equation}
 {\mathcal L}_{\rm{NC-NSI}} = -2\sqrt 2 \, G_F \,\varepsilon^{C\,f}_{\alpha\beta} 
 \, (\bar\nu_{\alpha}\gamma^\rho P_L\nu_{\beta}) \, 
 (\bar f \gamma_\rho P_C f),
\end{equation}
where $G_{F}$ is the Fermi coupling constant, $\varepsilon^{C\,f}_{\alpha
\beta}$ is the dimensionless parameter which represents the strength of NSI 
relative to $G_{F}$, and $\nu_\alpha$ and $\nu_\beta$ are the neutrino fields 
of flavor $\alpha$ and $\beta$ respectively. Here, $f$ denotes the matter fermions, electron ($e$), up-quark ($u$), and down-quark ($d$). 
Here, $P_L = (1 - \gamma_5)/2$, $P_R = (1 + \gamma_5)/2$, and the subscript $C = L,\,R$ 
expresses the chirality of the $ff$ current. 
Due to the hermiticity of the interaction, we have $\varepsilon^{C\,f}_{\alpha\beta} = (\varepsilon^{C\,f}_{\beta\alpha})^*$.

The NSI's of neutrino with matter fermions can give rise to the additional matter induced 
potentials apart from the standard MSW potential due to the $W$-mediated interactions in matter as denoted by $V_{CC}$.  
The total relative strength of the matter induced potential generated by the NC NSI's of neutrinos with all the matter fermions 
($\nu_\alpha + f \rightarrow \nu_\beta + f $) can be written in the following fashion,  
\begin{equation}
 \varepsilon_{\alpha\beta} = \sum_{f=e,u,d} \frac{V_f}{V_{CC}}(\varepsilon^{L\,f}_{\alpha\beta}+
 \varepsilon^{R\,f}_{\alpha\beta})\,\,,
\end{equation}
where $V_f$ = $\sqrt 2$ $G_{F}\, N_{f}$, $f=e, u, d$. The quantity  $N_f$ denotes 
the number density of matter fermion $f$ in the medium. For antineutrino, $V_f \rightarrow -V_f$ and $V_{CC} \rightarrow -V_{CC}$. In general, 
the total matter induced potential in presence of all the possible NC non-standard 
interactions of neutrino with matter fermions can be written as 
\begin{equation}
 H_{mat} = \sqrt 2 G_{F} N_{e} 
 \left( \begin{matrix}
1+\varepsilon_{ee}& \varepsilon_{e\mu} & \varepsilon_{e\tau}\\
\varepsilon_{e\mu}^* & \varepsilon_{\mu\mu} & \varepsilon_{\mu\tau}\\
\varepsilon_{e\tau}^* & \varepsilon_{\mu\tau}^* & \varepsilon_{\tau\tau}
\end{matrix}\right)  \,\,.
\label{eq:hmat-chapnsi}
\end{equation}
In the present study, we focus our investigation to flavor violating NSI parameter 
$\varepsilon_{\mu\tau}$, that is, we only allow $\varepsilon_{\mu\tau}$ to be non-zero 
in our analysis, and assume all other NSI parameters to be zero. We also consider 
$\varepsilon_{\mu\tau}$ to be real entertaining both of its negative and positive values. 
Since the atmospheric neutrino oscillation is mainly governed by $\nu_\mu\rightarrow\nu_\tau$ transition, it is expected 
that NSI parameter $\varepsilon_{\mu\tau}$ would have significant impact on this 
oscillation channel, which in turn can modify $\nu_\mu\rightarrow\nu_\mu$ survival 
probability by a considerable amount. We can study this effect by observing the atmospheric 
neutrinos at the proposed 50 kt magnetized ICAL detector. If we will not see any 
significant deviation from the standard $\mu^-$ and $\mu^+$ event spectra at ICAL, 
we can use this fact to place tight constraints on NSI parameter $\varepsilon_{\mu\tau}$. 
This is the main theme of our present study.

This article is organized in the following fashion. 
In Sec.\,\ref{sec:nsi-formalism-nsichap}, we briefly review the existing bounds 
on NSI parameter $\varepsilon_{\mu\tau}$ from various neutrino oscillation experiments. 
We discuss the possible modification in oscillation probabilities of neutrino and 
antineutrino due to non-zero $\varepsilon_{\mu\tau}$ in Sec.\,\ref{sec:osc-prob-nsichap}. 
In Sec.\,\ref{sec:event-plot-nsichap}, we present the expected total $\mu^-$ and $\mu^+$ 
events and their distributions as a function of reconstructed $E_\mu$ and $\cos\theta_\mu$ 
for the following three cases: (i) $\varepsilon_{\mu\tau} = 0$ (SM), (ii) $\varepsilon_{\mu\tau} 
= 0.05$, and (iii) $\varepsilon_{\mu\tau} = -0.05$ using 500 kt$\cdot$yr exposure of the ICAL 
detector. In Sec.\,\ref{sec:method-analysis-nsi}, we discuss the numerical procedure 
and various binning schemes that we use in our analysis. 
We present all the results of our study in Sec.\,\ref{sec:results-nsichap} where we show 
the following: (a) The possible improvement in the sensitivity of the ICAL detector 
in constraining $\varepsilon_{\mu\tau}$ due to the inclusion of events with $E_\mu$ in range of  
11 to 21 GeV in addition to the events that belong to the $E_\mu$ in range of 1 to 11\,GeV. (b) How much 
the limit on $\varepsilon_{\mu\tau}$ can be improved by considering the information on reconstructed 
hadron energy ($E'_{\rm had}$) as an additional observable along with reconstructed 
variables $E_\mu$ and $\cos\theta_\mu$ on an event-by-event basis. (c) We show the 
advantage of having charge identification (CID) capability in the ICAL detector in placing 
competitive constraint on $\varepsilon_{\mu\tau}$. (d) We present the expected limits on $\varepsilon_{\mu\tau}$ considering different exposures of the 50 kt ICAL detector. (e) We also explore the possible 
impact of non-zero $\varepsilon_{\mu\tau}$ in determining the mass hierarchy and in 
the precision measurement of atmospheric oscillation parameters. We provide a summary 
of this study in Sec.\,\ref{sec:conclusions-nsichap}.

\section{Existing Limits on NSI Parameter $\varepsilon_{\mu\tau}$}
\label{sec:nsi-formalism-nsichap}
There are existing constraints on the NSI parameter $\varepsilon_{\mu\tau}$ from various neutrino 
oscillation experiments. 
The Super-Kamiokande collaboration performed an analysis of the atmospheric neutrino data 
collected during its phase-I and -II run assuming only NSI's with $d$-quarks\,\cite{Mitsuka:2011ty}. 
The following bounds at $90\%$ C.L. are  obtained:
 \begin{equation}
  |\varepsilon| = |\varepsilon^d_{\mu\tau}| < {0.011} \,,\hspace{1 cm}  |\varepsilon'|  = |\varepsilon^{d}_{\tau\tau} - \varepsilon^{d}_{\mu\mu}| < 0.049\,.
  \end{equation}
Since $N_d = N_u = 3N_e$ for an electrically neutral and isoscalar Earth matter, the above 
constraints as obtained in Ref.\,\cite{Mitsuka:2011ty} are actually on the 
NSI parameters $\varepsilon_{\alpha\beta}/3$. Therefore, the above constraints 
at $90\%$ C.L. can be interpreted as
 \begin{equation}
  |\varepsilon_{\mu\tau}| < {0.033} \,, \hspace{1 cm} 
  |\varepsilon_{\tau\tau} - \varepsilon_{\mu\mu}| < 0.147\,.
  \end{equation}
Recently, the authors in Ref.\,\cite{Salvado:2016uqu} considered the possibility of NSI's in $\mu$-$\tau$ 
sector in the one-year high-energy through-going muon data of IceCube. In their analysis, they included 
various systematic uncertainties on both the atmospheric neutrino flux and detector properties, which 
they incorporated via several nuisance parameters. They obtained the following limits
\begin{equation}
-6.0\times 10^{-3}\,<\,\varepsilon_{\mu\tau}\,<\,5.4\times10^{-3} \, \hspace{.2cm} {\rm at \,\,\, 90\%\, 
credible\,interval\,(C.I.)}.
\end{equation}
The IceCube-DeepCore  collaboration also searched for NSI's involving $\varepsilon_{\mu\tau}$\,\cite{Aartsen:2017xtt}. 
Using their three years of atmospheric muon neutrino disappearance data, they placed the following constraint at 
$90\%$ confidence level  
\begin{equation}
-6.7\times 10^{-3}\,<\,\varepsilon_{\mu\tau}\,<\,8.1\times10^{-3}\,.
\end{equation}
A preliminary analysis to constrain the NSI parameters in context of the ICAL detector was performed 
in Ref.\,\cite{Choubey:2015xha}. Using an exposure of 500 kt$\cdot$yr and considering  only muon momentum 
($E_\mu, \cos\theta_\mu$) as 
observable, the authors in Ref.\,\cite{Choubey:2015xha} obtained the following bound 
\begin{equation}
 -0.015\,(-0.027)< \varepsilon_{\mu\tau} < 0.015\,(0.027)  \hspace{.2cm} {\rm at \,\,\, 90\,\,} (3\sigma)\,\, {\rm C.L.\,\, with \,\, NH}\,. 
\end{equation}
In the present study, we estimate new constraints on $\varepsilon_{\mu\tau}$ considering the reconstructed 
hadron energy ($E'_{\rm had}$) as an additional observable along with the reconstructed $E_\mu$ 
and $\cos\theta_\mu$
on an event-by-event basis at the ICAL detector. 

\section{$\nu_\mu\rightarrow\nu_\mu$ transition with non-zero $\varepsilon_{\mu\tau}$ }
\label{sec:osc-prob-nsichap}
This section is devoted to explore the effect of non-zero $\varepsilon_{\mu\tau}$ in the oscillation 
of atmospheric neutrino and antineutrino propagating long distances through the Earth matter. For this, 
we numerically estimate the three-flavor oscillation probabilities including NSI parameter 
$\varepsilon_{\mu\tau}$ and using the PREM profile\, \cite{Dziewonski:1981xy} for the Earth matter density. 
The NSI parameter $\varepsilon_{\mu\tau}$ modifies the evolution of neutrino in matter, which in the flavor 
basis takes the following form, 
 \begin{equation}
  {\rm i} \frac{\rm{d}}{\rm{dt}} \left( \begin{matrix} \nu_e(t) \\ \nu_\mu(t) \\ \nu_\tau(t) \end{matrix} \right ) 
  = \frac{1}{2E} \, \left[ U \left( \begin{array}{ccc} 0 & 0 & 0 \\ 0 & \Delta m^2_{21} & 0 \\ 0 & 0 & 
  \Delta m^2_{31} \end{array} \right) U^\dag
  + 2\sqrt 2 G_{F} N_{e} E \, 
 \left( \begin{matrix}
1  & 0 & 0  \\
0 & 0 & \varepsilon_{\mu\tau}\\
0 & \varepsilon_{\mu\tau} & 0 
\end{matrix}\right) \right] \left( \begin{matrix} \nu_e \\ \nu_\mu \\ \nu_\tau \end{matrix} \right )\,\,\,, 
 \end{equation}
 where  $\varepsilon_{\mu\tau}$  is real in our analysis. 
\begin{figure}[htb!]
\begin{minipage}{.32\linewidth}
  \includegraphics[width=\textwidth]{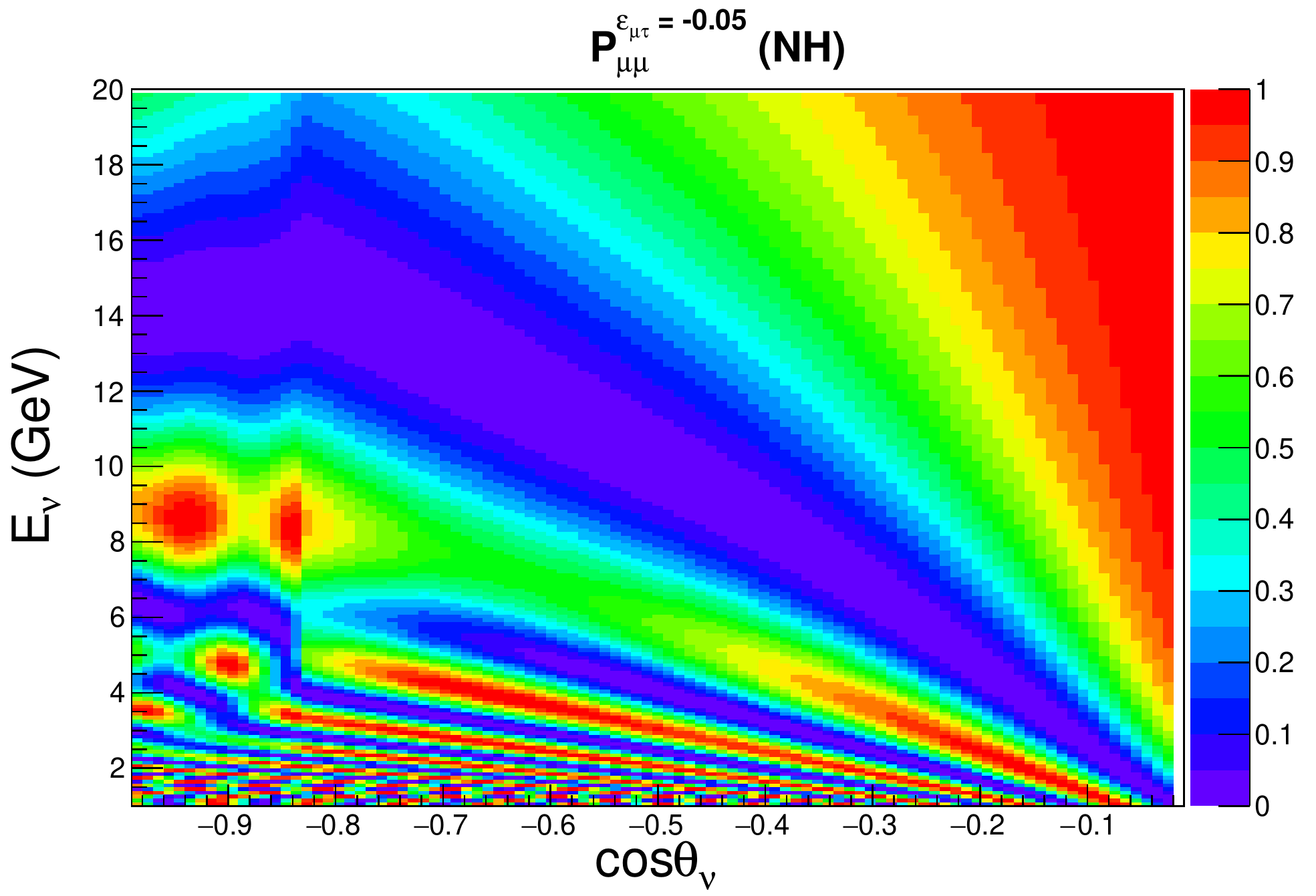}
  \end{minipage}
 \hfill
  \begin{minipage}{.32\linewidth}
   \includegraphics[width=\textwidth]{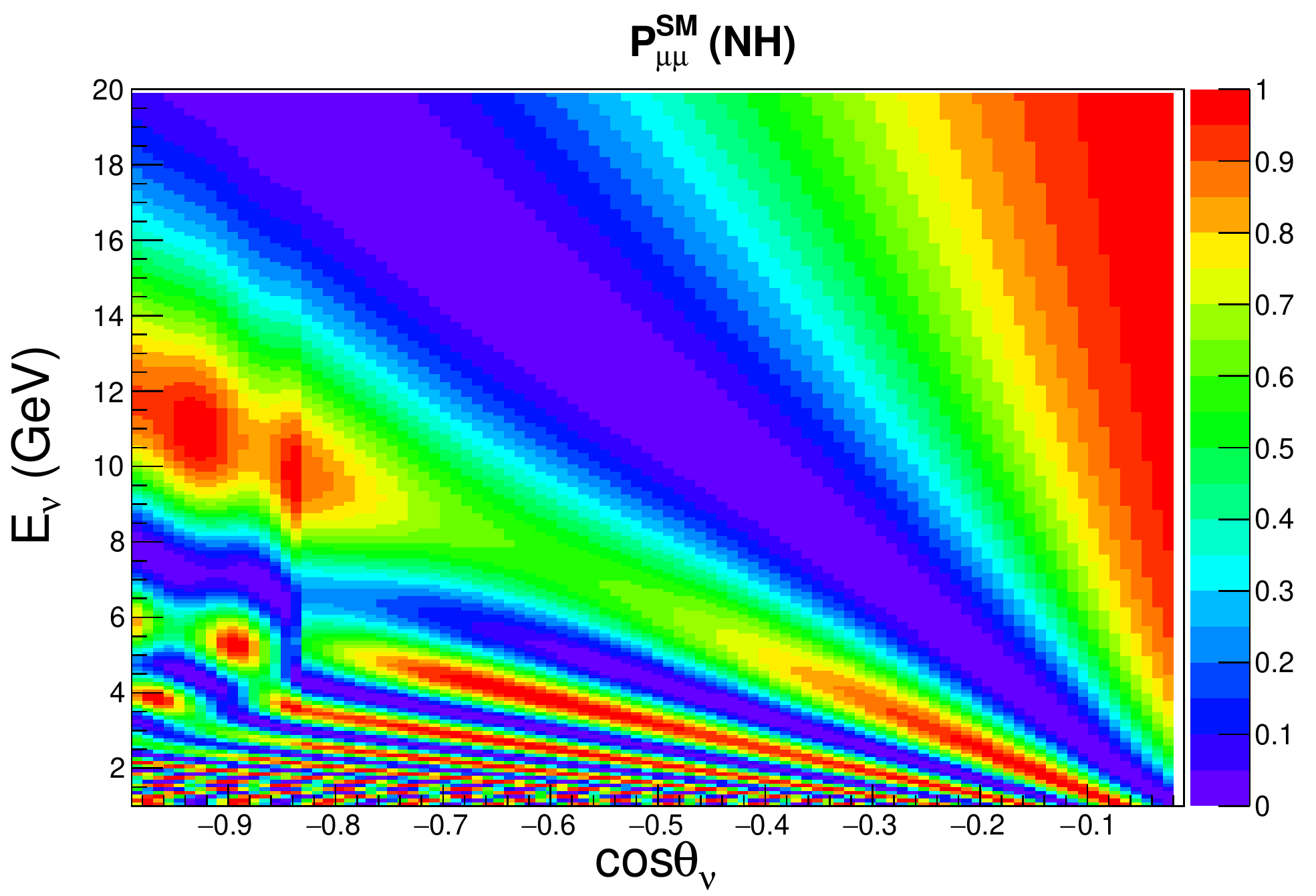}
  \end{minipage}
\hfill
 \begin{minipage}{.32\linewidth}
  \includegraphics[width=\textwidth]{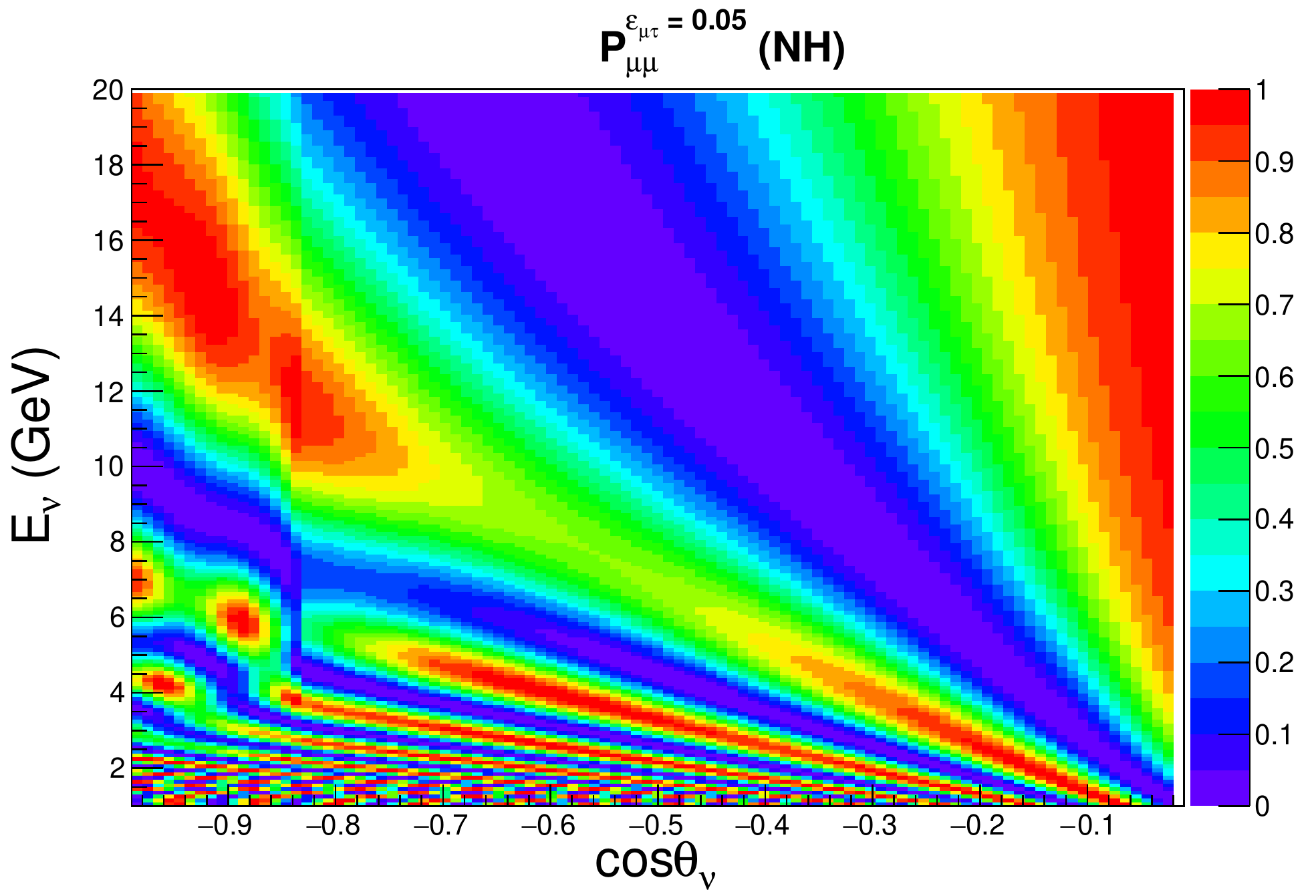}
  \end{minipage}
 \hfill
  \begin{minipage}{.32\linewidth}
   \includegraphics[width=\textwidth]{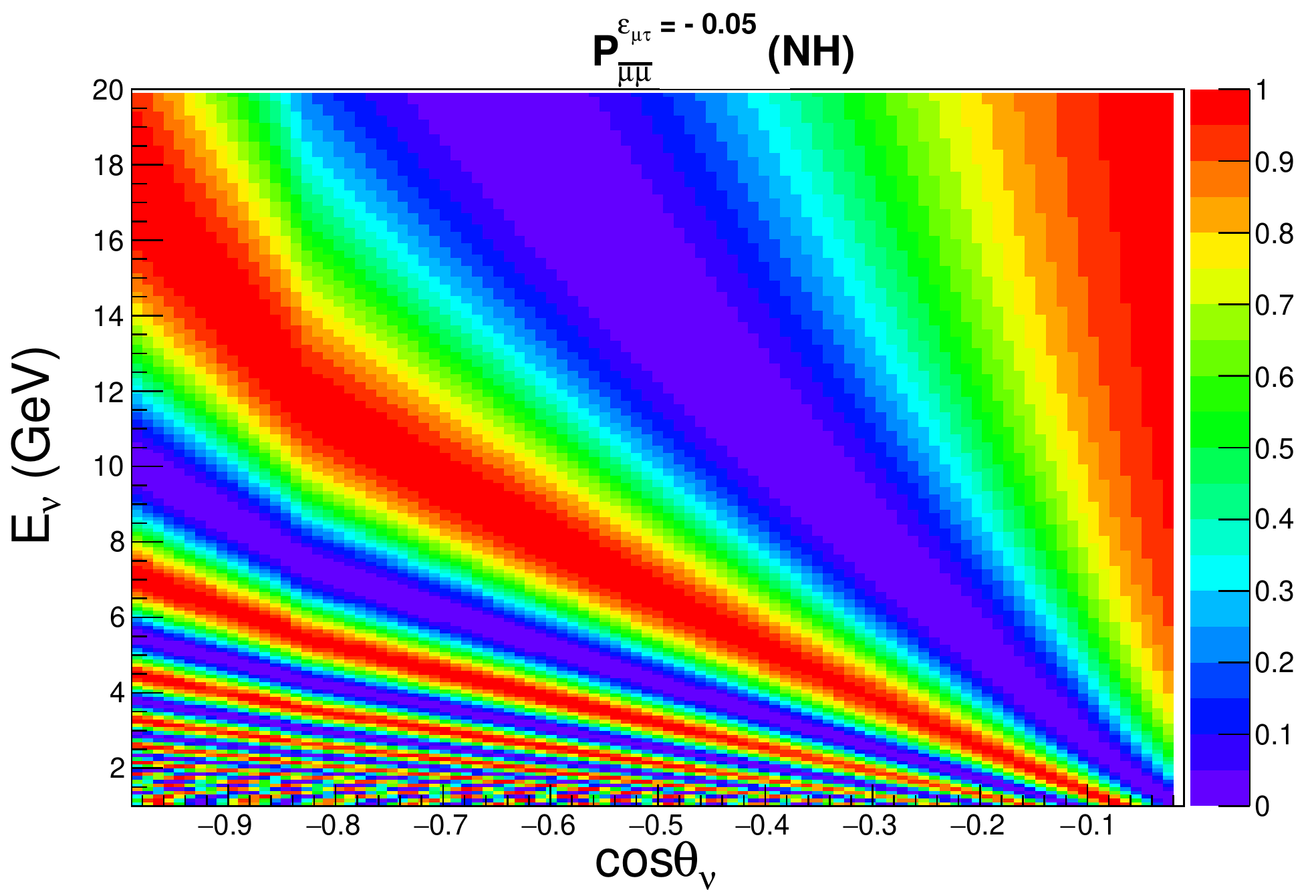}
  \end{minipage}
\hfill
 \begin{minipage}{.32\linewidth}
  \includegraphics[width=\textwidth]{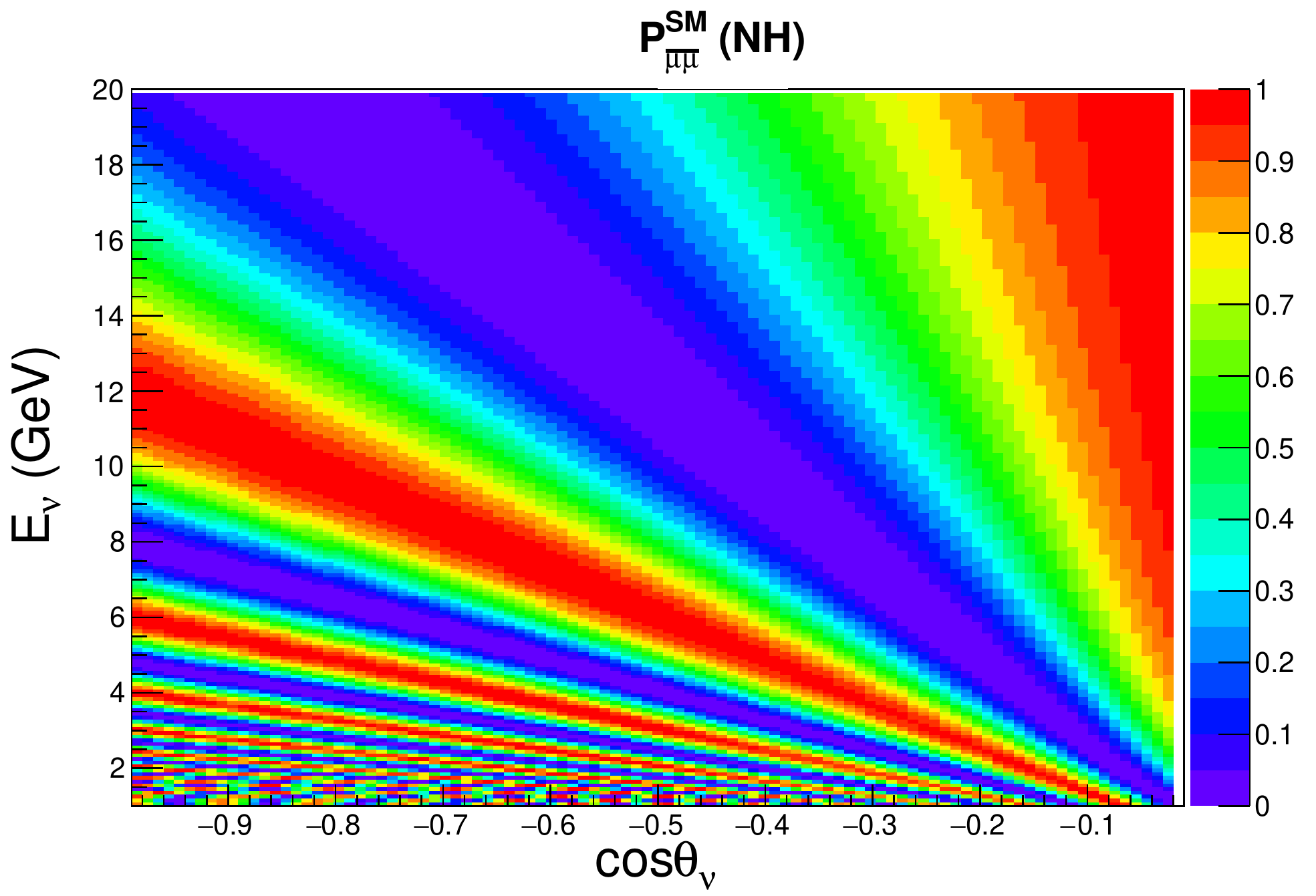}
  \end{minipage}
 \hfill
  \begin{minipage}{.32\linewidth}
   \includegraphics[width=\textwidth]{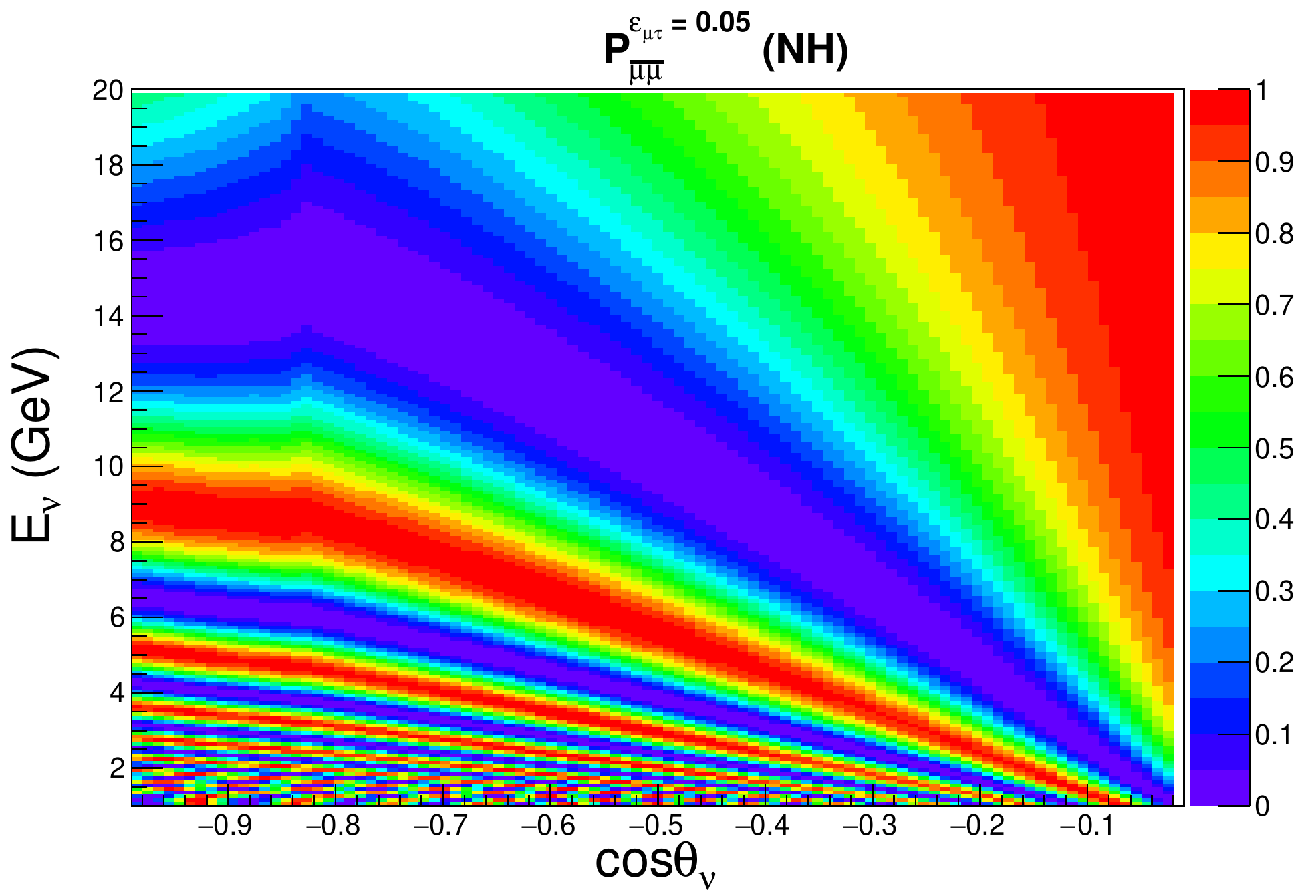}
  \end{minipage}
  \mycaption{The oscillograms for $\nu_\mu\rightarrow\nu_\mu$ ($\bar\nu_\mu\rightarrow\bar\nu_\mu$) channel 
  in  $E_\nu$, $\cos\theta_{\nu}$ plane are shown in top (bottom) panels for three different scenarios: 
  i) $\varepsilon_{\mu\tau} = -0.05$ (left panel), ii) $\varepsilon_{\mu\tau}=0.0 $ (the SM case, middle panel), 
  and iii) $\varepsilon_{\mu\tau} = 0.05$  (right panel). Here, in all the panels, we assume NH.}
  \label{fig:osc-sm-nsi}
  \end{figure}

In upper panels of Fig.\,\ref{fig:osc-sm-nsi}, we present the oscillograms for $\nu_\mu$ survival channel in 
the plane of $\cos\theta_\nu$  vs. $E_\nu$ considering NH. Here, we draw the oscillograms for three different 
cases:  i) $\varepsilon_{\mu\tau} = -0.05$ (left panel), ii) $\varepsilon_{\mu\tau}=0.0 $ (the SM case, middle 
panel), and iii) $\varepsilon_{\mu\tau} = 0.05$  (right panel). The lower panel depicts the same but for  
$\bar\nu_\mu\rightarrow\bar\nu_\mu$ oscillation channel.  
To prepare Fig.\,\ref{fig:osc-sm-nsi}, we take the following benchmark values of vacuum oscillation parameters 
in three-flavor framework:
$\sin^2\theta_{23} =0.5$, $\sin^2 2\theta_{13} =0.1$, $\sin^2 \theta_{12} =0.3$, $\Delta m^2_{21} = 7.5 \times 
10^{-5}$ eV$^2$, $\Delta m_{\rm eff}^{2} = 2.4 \times 10^{-3}$ eV$^2$, and $\delta_{\textrm{CP}} =0^\circ$. 
We estimate the value of $\Delta m^2_{31}$ from $\Delta m_{\rm eff}^{2}$\footnote{The effective mass splitting is related to $\Delta m^{2}_{31}$ as follows\,\cite{Nunokawa:2005nx,deGouvea:2005hk} 
 \begin{equation}
  \Delta m_{\textrm{eff}}^{2} = \Delta m^{2}_{31}\,-\Delta m^{2}_{21}\left(\cos^{2}\theta_{12}\,
				  -\,\cos\delta_{\rm CP}\, \sin\theta_{13}\,\sin 2\theta_{12}\,\tan\theta_{23}\right)\,.
\label{eq:dmseff-chap3}				  
 \end{equation}}, where $\Delta 
m_{\textrm{eff}}^{2}$ has the same magnitude for NH and IH with positive and negative signs respectively.    
It is evident from the upper panels of Fig.\,\ref{fig:osc-sm-nsi} that in the presence of negative (see left panel) and positive (see right panel) non-zero values of $\varepsilon_{\mu\tau}$,  
$\nu_\mu$ survival probabilities get modified substantially at higher energies and longer baselines, where vacuum oscillation dominates. We observe the similar changes in case of $\bar\nu_\mu$ oscillation probabilities as well (see lower panels). 

Another broad feature which has been emerging from Fig.\,\ref{fig:osc-sm-nsi} is that the $\nu_\mu \rightarrow \nu_\mu $ oscillation probabilities with positive (negative) $\varepsilon_{\mu\tau}$ as shown in upper right (left) panel are similar to that of $\bar\nu_\mu \rightarrow \bar\nu_\mu$ transition probabilities with negative (positive) $\varepsilon_{\mu\tau}$ as can be seen from lower left (right) panel. We can understand this behaviour with the help of following approximate analytical expression of  $\nu_\mu\rightarrow\nu_\mu$ transition probability. Assuming $\Delta m^2_{21} L/4E \rightarrow 0$ and $\theta_{13} = 0$,  $\nu_\mu\rightarrow\nu_\mu$ oscillation channel in the presence of non-zero $\varepsilon_{\mu\tau}$ and under the constant matter density approximation takes the form\,\cite{Gago:2001xg,GonzalezGarcia:2004wg} 
\begin{equation}
 P_{\nu_\mu\rightarrow\nu_\mu} = 1 - \sin^2 2\theta_{\rm eff} \, \sin^2\left[ \xi \,\frac{\Delta m^2_{31} L}{4 E} \right]\,,
 \label{eq:pmumu-nsi-mutau-omsd}
\end{equation}
where
\begin{equation}
 \sin^2 2\theta_{\rm eff} = \frac{|\sin 2\theta_{23} \pm 2 \eta_{\mu\tau}|^2}{\xi^2}\,,
 \label{eq:pmumu-thetaeff-omsd}
\end{equation}
\begin{equation}
 \xi = \sqrt{|\sin 2\theta_{23} \pm 2 \eta_{\mu\tau}|^2 + \cos^2 2\theta_{23}}\,,
 \label{eq:pmumu-xi-omsd}
\end{equation}
and
\begin{equation}
 \eta_{\mu\tau} = \frac{ 2 E \,V_{CC} \,\varepsilon_{\mu\tau}}{\Delta m^2_{31}}\,.
\end{equation}
In Eqs.\,\ref{eq:pmumu-thetaeff-omsd} and \ref{eq:pmumu-xi-omsd}, positive and negative signs in front of $\eta_{\mu\tau}$ are associated with the normal and inverted hierarchy respectively.
In case of maximal mixing ($\theta_{23} = 45^{\circ}$), Eq.\,\ref{eq:pmumu-nsi-mutau-omsd} boils down to the following simple expression\,\cite{Hollander:2014iha}
\begin{equation}
 P_{\nu_\mu\rightarrow\nu_\mu} = \cos^2 \left[L\left( \frac{\Delta m^2_{31}}{4E} + \varepsilon_{\mu\tau} V_{CC} \right)\right]\,.
 \label{eq:pmumu-nu-final-omsd}
\end{equation}
Since for antineutrino, $V_{CC} \rightarrow -V_{CC}$,  following Eq.\,\ref{eq:pmumu-nu-final-omsd}, we can write    
\begin{equation}
 P_{\nu_\mu\rightarrow \nu_\mu} (-\varepsilon_{\mu\tau}) = P_{\bar\nu_\mu \rightarrow \bar\nu_\mu} (\varepsilon_{\mu\tau})\, 
 \label{eq2}
\end{equation} 
and 
\begin{equation}
 P_{\nu_\mu\rightarrow \nu_\mu} (\varepsilon_{\mu\tau}) = P_{\bar\nu_\mu \rightarrow \bar\nu_\mu} (-\varepsilon_{\mu\tau})\,.
 \label{eq1}
\end{equation}   
Eq.\,\ref{eq2} and Eq.\,\ref{eq1} explain the broad features in Fig.\,\ref{fig:osc-sm-nsi} that we mention above. 
\begin{figure}[t!]
 \begin{minipage}{.49\linewidth}
  \includegraphics[width=\textwidth]{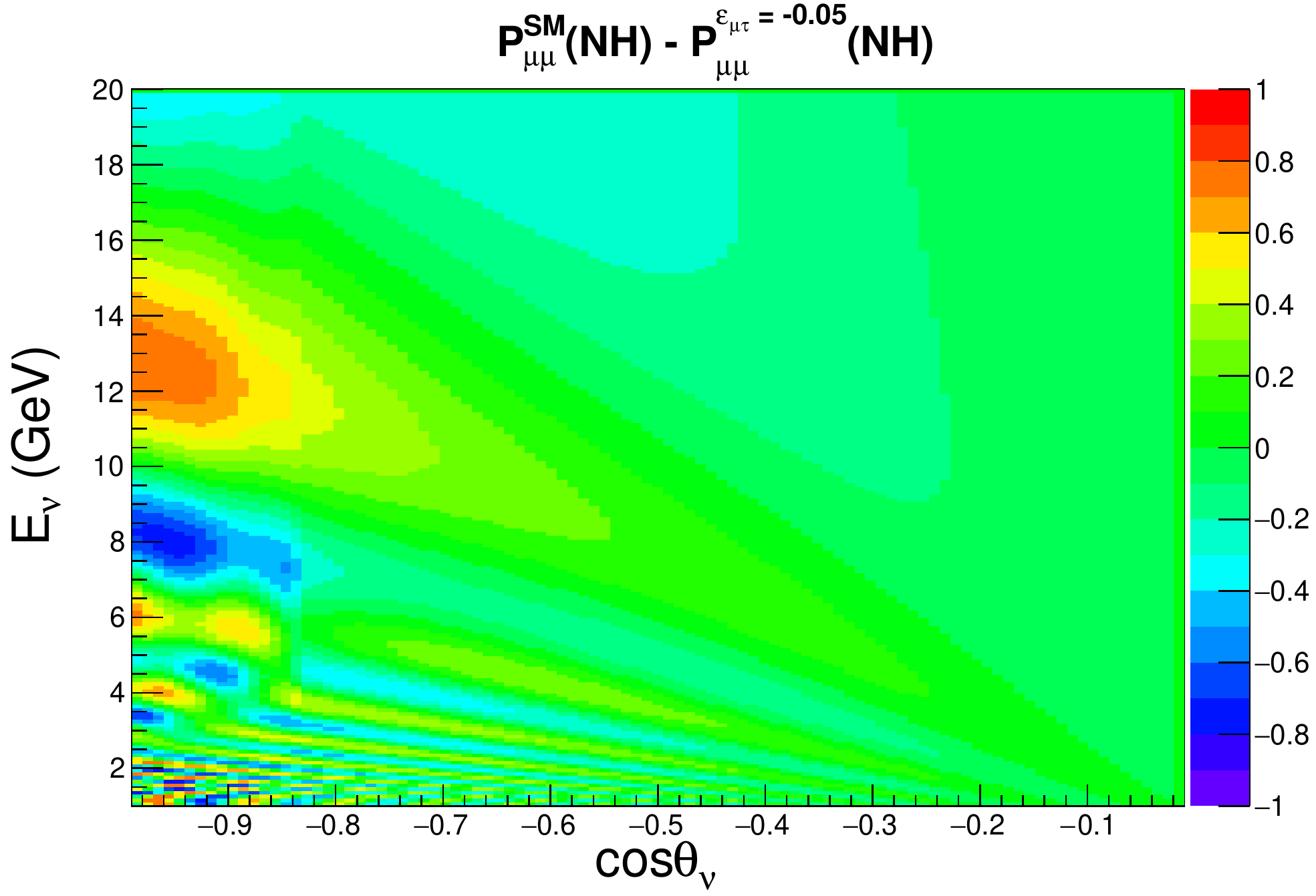}
  \end{minipage}
 \hfill
  \begin{minipage}{.49\linewidth}
   \includegraphics[width=\textwidth]{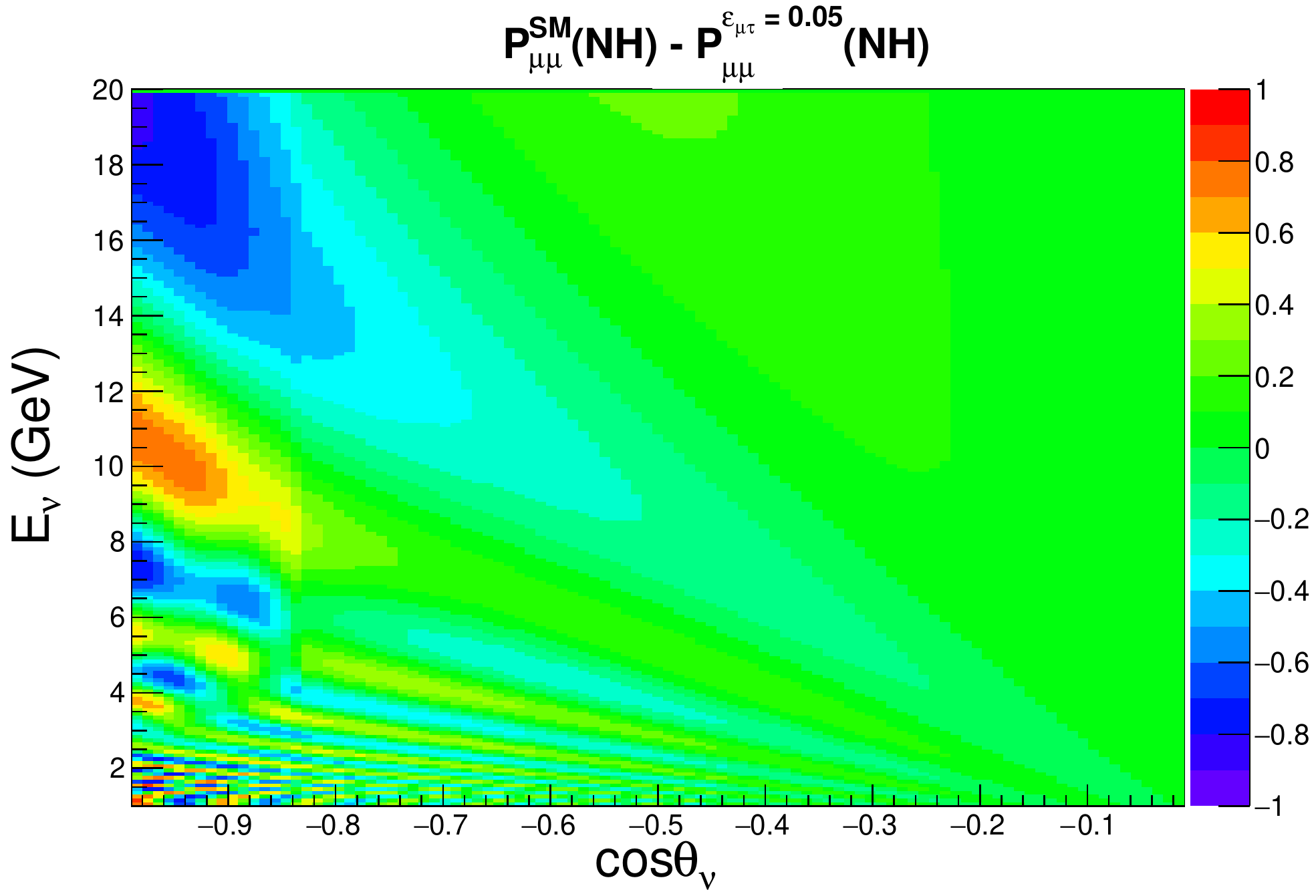}
  \end{minipage}
\hfill
 \begin{minipage}{.49\linewidth}
  \includegraphics[width=\textwidth]{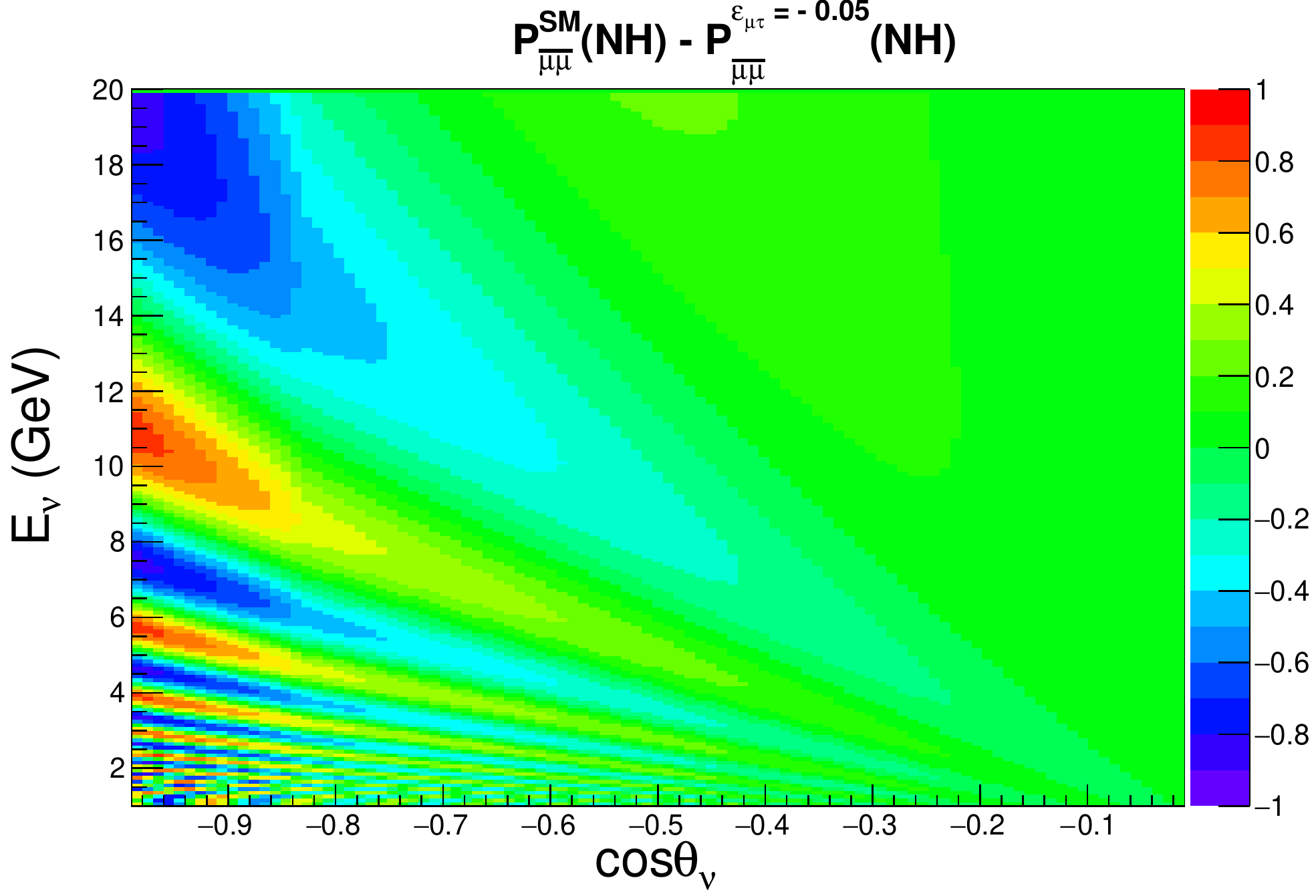}
  \end{minipage}
 \hfill
  \begin{minipage}{.49\linewidth}
   \includegraphics[width=\textwidth]{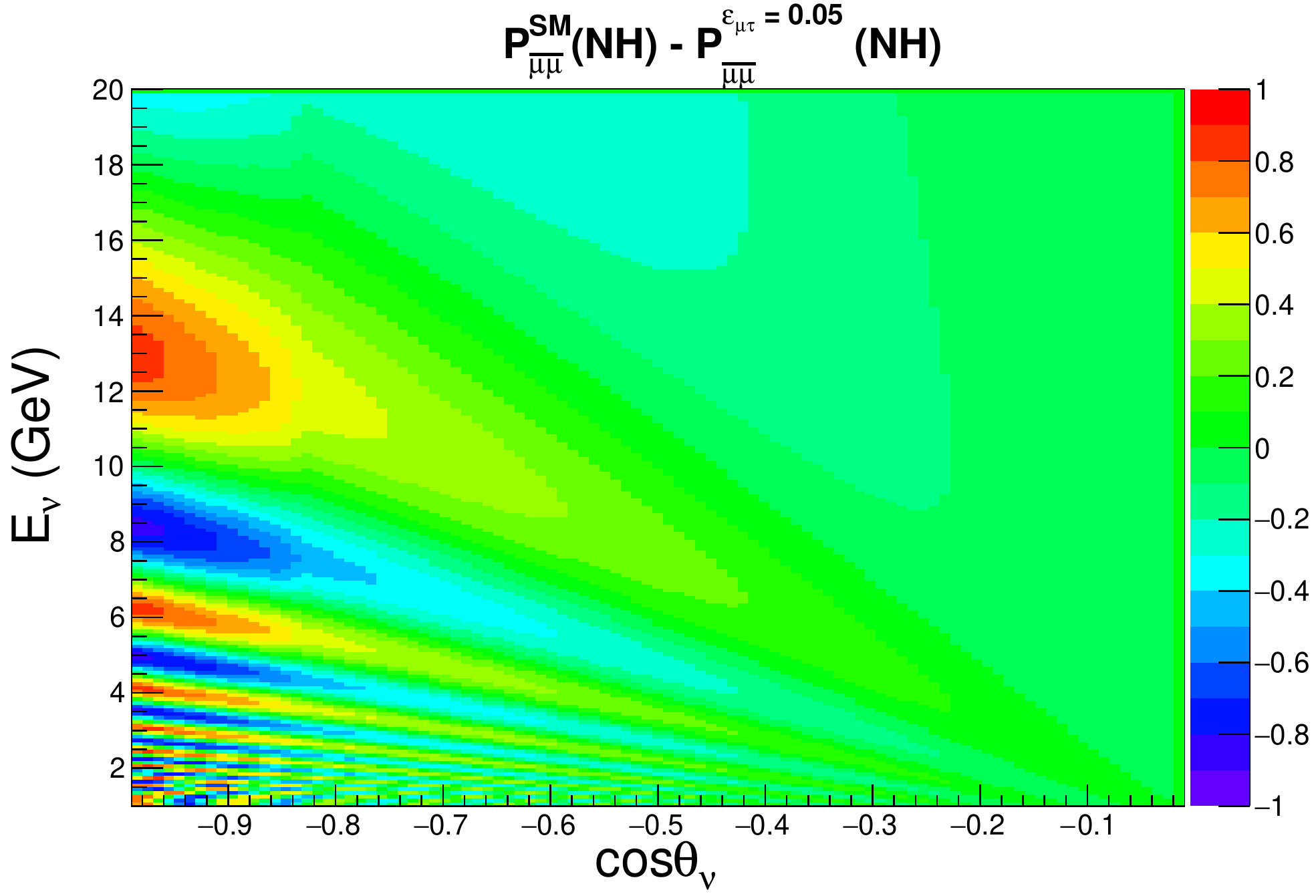}
  \end{minipage}
  \mycaption{The upper left panel shows the difference in $\nu_\mu\rightarrow\nu_\mu$ oscillation channel 
  between the SM case ($\varepsilon_{\mu\tau}= 0$) and $\varepsilon_{\mu\tau}= -0.05$. In the top right panel, 
  the difference is due to the SM case and $\varepsilon_{\mu\tau}= 0.05$. The lower panels are for $\bar\nu_\mu
  \rightarrow\bar\nu_\mu$ oscillation channel. 
  Here, in all the panels, we assume NH. }
  \label{fig:osc-dif}
  \end{figure} 

 To have a better look at the changes induced by non-zero $\varepsilon_{\mu\tau}$ as compared to the SM case, we give Fig.\,\ref{fig:osc-dif} 
where we plot the difference in $\nu_\mu\rightarrow\nu_\mu$ survival channel considering the cases 
$\varepsilon_{\mu\tau} = 0$ (the SM case) and $\varepsilon_{\mu\tau} = -0.05$ (see top left panel). In 
top right panel, we present the same for the cases of $\varepsilon_{\mu\tau} = 0$ (the SM case) and 
$\varepsilon_{\mu\tau} = 0.05$. The lower panels are for antineutrinos. In all the panels, we see a 
visible difference in $\nu_\mu$ survival channel due to the presence of non-zero $\varepsilon_{\mu\tau}$ as 
compared to the SM case ($\varepsilon_{\mu\tau} = 0.0$) at higher baselines with $\cos\theta_\nu$ in the 
range $-1$ to $-0.8$. This range of $\cos\theta_\nu$ corresponds to the baseline in the range $\sim$ 12700 km
to 10000 km where neutrino and antineutrino  mostly travel through inner and outer part of the Earth's core\footnote{According to a simplified 
version of the PREM profile\,\cite{Dziewonski:1981xy}, the inner core has a radius of $\sim$ 1220 km 
with an average density of 13 g/cm$^3$. For outer core, $R_{\rm min} \simeq 1220$ km and $R_{\rm max} \simeq$ 
3480 km with an average density of 11.3 g/cm$^3$. Note that in our analysis, we consider the detailed 
version of the PREM. } and have access to large Earth matter effect. Also, we see a trend that the impact 
of NSI's is large at higher energies where the three-flavor oscillations are suppressed because the 
oscillation lengths  ($L^{\rm osc} =  \frac{4 \pi E}{ \Delta m^2_{ij} } $) are large at higher energies.

To have a quantitative estimate of the difference in $\nu_\mu\rightarrow \nu_\mu$ oscillation channel due to non-zero $\varepsilon_{\mu\tau}$  as compared to $\varepsilon_{\mu\tau} =0$ case, we use  
Eq.\,\ref{eq:pmumu-nu-final-omsd} and obtain the following expression 
\begin{equation}
 P^{\rm diff}_{\mu\mu} \equiv P^{\rm SM}_{\mu\mu} - P^{\varepsilon_{\mu\tau}}_{\mu\mu} = \sin \left( V_{CC}\,\varepsilon_{\mu\tau} \,L \right)\,\sin\left[L\left(\frac{\Delta m^2_{31}}{2E} + V_{CC}\,\varepsilon_{\mu\tau}\right) \right] \,.
\label{eq:diff-pmumu}
 \end{equation}
Eq.\,\ref{eq:diff-pmumu} clearly suggests that the impact of NSI is proportional to both NSI induced matter potential ($V_{CC} \varepsilon_{\mu\tau}$) and baseline ($L$). 
We see in upper left and lower right panels of Fig.\,\ref{fig:osc-dif} that around $E\sim 18$ GeV and $\cos\theta_\nu\sim -\, 0.9$, $ P^{\rm diff}_{\mu\mu}$ approaches to zero suggesting that the impact of NSI is negligible. We observe the opposite behaviour in upper right and lower left panels, where around $E\sim 18$ GeV and $\cos\theta_\nu\sim -\,0.9$, $ P^{\rm diff}_{\mu\mu}$ attains a quite large value of $-\, 0.8$ suggesting that the influence of NSI is significant there.  
Here, $\cos\theta_\nu = -\,0.9$ corresponds to $L = 11500$ km for which the line-averaged constant Earth matter density according to the PREM\,\cite{Dziewonski:1981xy} profile is 6.8 g/cm$^3$. 
Therefore, the standard line-averaged Earth matter potential\footnote{The standard neutrino matter potential due to the $W$-mediated interactions with the ambient electrons can be written as a function of matter density $\rho$ as follows:\begin{equation}
V_{CC} \simeq 7.6  \,\times\, Y_e \,\times \,\frac{\rho}{\rm 10^{14} g/cm^{3}} \,\,\, {\rm eV}\,, 
\label{eq:vcc-chap2}
\end{equation}
where $Y_e$ ($\frac{N_e}{N_p + N_n}$) is the relative number density. For electrically,  
neutral and isoscalar medium, $N_e = N_p$ $= N_n$, and therefore, $Y_e =0.5$.                                                                                                                                                                                                                                                                                                                     } for 11500 km baseline is $V_{CC}\sim 2.6\times 10^{-13}$ eV. Considering $\varepsilon_{\mu\tau} = 0.05$,  we get 
\begin{equation}
V_{CC} \,\varepsilon_{\mu\tau} \,L = 2.6\times 10^{-13} \,{\rm eV} \,\times 0.05 \times 11500 \times 5.06 \times 10^9 ({\rm eV})^{-1}\sim 0.76 \,. 
\end{equation}
For $E = 18$\,GeV, $L =11500$ km, and $\Delta m^2_{31} = 2.36\times 10^{-3}$ eV$^2$ (this mass-squared difference is obtained from Eq.\,\ref{eq:dmseff-chap3} using benchmark values of oscillation parameters),  
\begin{equation}
\Delta m^2_{31} L/2E = 3.8 \,. 
\end{equation}
Thus, for $E = 18$ GeV and $L = 11500$ km, from Eq.\,\ref{eq:diff-pmumu}, we obtain 
\begin{equation}
  P^{\rm diff}_{\mu\mu}(\varepsilon_{\mu\tau} = -0.05) =  P^{\rm diff}_{\bar\mu\bar\mu} (\varepsilon_{\mu\tau} = 0.05) = \sin (3.8  - 0.76 ) \, \sin (0.76 ) \sim 0.06\,
  \label{eq3}
\end{equation}
and 
\begin{equation}
 P^{\rm diff}_{\mu\mu}(\varepsilon_{\mu\tau} = 0.05) =  P^{\rm diff}_{\bar\mu\bar\mu} (\varepsilon_{\mu\tau} = -0.05) = \sin (3.8  + 0.76 ) \,\sin (0.76 ) \sim -0.7\,.
 \label{eq4}
\end{equation}
Eq.\,\ref{eq3} and Eq.\,\ref{eq4} confirm the observations regarding $P^{\rm diff}_{\mu\mu}$ and $P^{\rm diff}_{\bar\mu\bar\mu}$ in Fig.\,\ref{fig:osc-dif} that we mention above. We know that due to its CID  capability, ICAL has an edge to resolve the issue of neutrino mass hierarchy (sign of $\Delta m^2_{31}$) by observing the Earth matter effect in $\mu^-$ and $\mu^+$ events separately\,\cite{Kumar:2017sdq}. Similarly, the four panels in Fig.\,\ref{fig:osc-dif} suggest that the CID capability of ICAL can provide useful information to determine the sign of NSI parameter $\varepsilon_{\mu\tau}$ for a particular choice of mass hierarchy.

\section{Expected Events at ICAL with non-zero $\varepsilon_{\mu\tau}$}
\label{sec:event-plot-nsichap}
The Monte Carlo based neutrino event generator NUANCE\,\cite{Casper:2002sd} is used to simulate the 
CC interactions of $\nu_\mu$  and $\bar\nu_\mu$ in the ICAL detector. To generate events 
in  NUANCE, we give a simple geometry of the ICAL detector with 150 alternate layers of iron 
and glass plates in each module. We have three such modules to account for the 50 kt ICAL 
detector. As far as the neutrino flux is concerned in generating the neutrino events in the 
present study, we use the flux as predicted at Kamioka\footnote{Preliminary calculation of the 
expected fluxes at the INO site has been performed in Ref.\,\cite{Athar:2012it,Honda:2015fha}. 
The visible differences between the neutrino fluxes at the Kamioka and INO sites appear at lower 
energies.  The main reason behind this is that the horizontal components of the geo-magnetic 
field are different at the Kamioka (30$\mu$T) and INO (40 $\mu$T) locations. We plan to use 
these new fluxes estimated for the INO site (see Ref.\,\cite{Honda:2015fha}) in future 
studies.}\,\cite{Honda:2011nf}. To reduce the statistical fluctuation, we generate the unoscillated 
CC neutrino and antineutrino events considering a very high exposure of 1000 years and 50 kt ICAL.   
Then, we implement various oscillation probabilities using the reweighting algorithm. Next, we fold the oscillated events with detector response for 
muon and hadron as described in Ref.\,\cite{Chatterjee:2014vta,Devi:2013wxa}.
In the present study, we assume that the ICAL particle reconstruction algorithms can separate the hits 
due to the hadron shower from the hits originating from a muon track with 100$\%$ efficiency. It means 
that whenever a muon is reconstructed, we consider all the other hits to be part of the hadronic shower
in order to calibrate the hadron energy. It also implies that the neutrino event reconstruction efficiency 
is same as the muon reconstruction efficiency. Finally, the reconstructed $\mu^-$ and $\mu^+$ events are 
scaled down to the exposure of 10 years for 50 kt ICAL. Now, we present the expected $\mu^-$ and $\mu^+$ 
events for 500 kt$\cdot$yr exposure of the ICAL detector assuming the SM case ($\varepsilon_{\mu\tau} = 0$)
and $\varepsilon_{\mu\tau} = \pm\, 0.05$.  To estimate these event rates, we use the values of oscillation 
parameters as considered in Sec.\,\ref{sec:osc-prob-nsichap} to draw the oscillograms. 

\subsection{Total Event Rates}
First, we address the following question: can we see the signature of non-zero 
$\varepsilon_{\mu\tau}$ in the total number of $\mu^-$ and $\mu^+$ events 
which will be collected at the ICAL detector over 10 years of running? To have 
an answer of this question, we estimate the total number of events for the 
following three cases: i) $\varepsilon_{\mu\tau} = 0.05$, ii) $\varepsilon_{\mu\tau} 
=0$ (the SM case), and iii) $\varepsilon_{\mu\tau} = -0.05$. We present these 
numbers in Table\,\ref{tab:tot-event} with NH and using 500 kt$\cdot$yr 
exposure of the ICAL detector integrating over entire ranges of $E_\mu$, 
$\cos\theta_\mu$, and $E'_{\rm had}$ that we consider in our analysis. 
\begin{table}[htb!]
 \begin{center}
 \begin{tabular}{| c| c| c|| c|c|}
   \hline\hline
    & \multicolumn{2}{c||}{low-energy (LE)} & \multicolumn{2}{c|}{high-energy (HE)} \\
\hline
   $\varepsilon_{\mu\tau}$ & $\mu^-$ & $\mu^+$  &  $\mu^-$ & $\mu^+$ \\
   \hline
   0.05 & \makecell[c]{ 4574 (total) \\4474  ($P_{\mu\mu}$) \\ 100 ($P_{ e \mu}$)}	
  & \makecell[c]{ 2029 (total) \\ 2016 ($P_{\bar\mu\bar\mu}$) \\ 13 ($P_{\bar e\bar\mu}$) } & 
  \makecell[c]{4879 (total) \\ 4778 ($P_{\mu\mu}$) \\ 101 ($P_{e\mu}$)} & 
  \makecell[c]{2192 (total) \\ 2179 ($P_{\bar\mu\bar\mu}$) \\ 13 ($P_{\bar e\bar\mu}$)}\\
  \hline
  SM & \makecell[c]{ 4562 (total) \\  4458 ($P_{\mu\mu}$) \\104 ($P_{e\mu}$)} &
  \makecell[c]{ 2035 (total) \\ 2022 ($P_{\bar\mu\bar\mu}$) \\ 13 ($P_{\bar e\bar\mu}$) } &
    \makecell[c]{ 4870 (total) \\ 4765 ($P_{\mu\mu}$) \\ 105 ($P_{e\mu}$)} & 
  \makecell[c]{ 2188 (total) \\ 2175 ($P_{\bar\mu\bar\mu}$) \\ 13 ($P_{\bar e\bar\mu}$)}\\
  \hline
 -0.05 & \makecell[c]{ 4553 (total) \\  4444 ($P_{\mu\mu}$) \\  109 ($P_{e\mu}$)} &
  \makecell[c]{ 2037 (total) \\ 2024 ($P_{\bar\mu\bar\mu}$) \\ 13 ($P_{\bar e\bar\mu}$) } &
    \makecell[c]{4890 (total) \\4780 ($P_{\mu\mu}$) \\ 110 ($P_{ e \mu}$)} & 
  \makecell[c]{2191 (total) \\ 2178 ($P_{\bar\mu\bar\mu}$) \\ 13 ($P_{\bar e\bar\mu}$)}\\
  \hline\hline
 \end{tabular}
 \end{center}
 \mycaption{ Expected number of $\mu^-$ and $\mu^+$ events for 500 kt$\cdot$yr exposure of the 
 ICAL detector considering low-energy (LE) and high-energy (HE) binning schemes. We present the 
 event rates for the following three cases: i) $\varepsilon_{\mu\tau} = 0.05$, ii)  
 $\varepsilon_{\mu\tau} =0$ (the SM case), and iii) $\varepsilon_{\mu\tau} = -0.05$. 
 Apart from showing the total $\mu^-$ event rates, we also give the estimates of 
 individual event rates  coming from $\nu_\mu\rightarrow\nu_\mu$ ($P_{\mu\mu}$) disappearance
 channel and $\nu_e\rightarrow\nu_\mu$ ($P_{e\mu}$) appearance channel. For $\mu^+$ events also,  
 we separately show the contributions from  $\bar\nu_\mu\rightarrow\bar\nu_\mu$ ($P_{\bar\mu\bar\mu}$) disappearance
 channel and $\bar\nu_e\rightarrow\bar\nu_\mu$ ($P_{\bar e\bar\mu}$) appearance channel. Here, 
 we consider NH and assume the benchmark values of the oscillation parameters as mentioned in 
 Sec.\,\ref{sec:event-plot-nsichap}.
 }
 \label{tab:tot-event}
\end{table}
As far as the binning schemes are concerned, we use the low-energy (LE) and high-energy (HE)  
binning schemes\footnote{For a detailed description of the two binning schemes 
that we consider in our analysis, see Sec.\,\ref{sec:chap-nsi-binning}.}, and for both 
these binning schemes, we take the entire range of $\cos\theta_\mu$ spanning over -1 to 1. 
The energy ranges for reconstructed $E_\mu$ and $E'_{\rm had}$ are different 
in LE and HE binning schemes. For LE binning scheme, $E_\mu \in $ [1,\,11]\,GeV and $E'_{\rm had} \in$ [0,\,15]\,GeV. 
In case of HE binning scheme, $E_\mu \in $ [1,\,21]\,GeV, and $E'_{\rm had} \in$ [0,\,25]\,GeV. 
When we increase the reconstructed muon energy from 11 GeV to 21 GeV and reconstructed 
hadron energy from 15\,GeV to 25\,GeV, the number of $\mu^-$ and $\mu^+$ events get 
increased by 300 and 150 respectively for 500 kt$\cdot$yr exposure of the ICAL detector. 
Apart from showing the total $\mu^-$ event rates in Table\,\ref{tab:tot-event}, we also present the estimates of 
individual events coming from $\nu_\mu\rightarrow\nu_\mu$ ($P_{\mu\mu}$) disappearance
channel and $\nu_e\rightarrow\nu_\mu$ ($P_{e\mu}$) appearance channel. Also, for $\mu^+$ events, 
we separately show the contributions originating from  $\bar\nu_\mu\rightarrow\bar\nu_\mu$ 
($P_{\bar\mu\bar\mu}$) disappearance and $\bar\nu_e\rightarrow\bar\nu_\mu$ ($P_{\bar e\bar\mu}$) 
appearance channels. Here, we see that only $\sim 2 \%$ of the total $\mu^-$ events at the ICAL 
detector come via the appearance channel.  
Note that the differences in the total number of $\mu^-$ and $\mu^+$ events between the SM case 
($\varepsilon_{\mu\tau} = 0$) and non-zero $\varepsilon_{\mu\tau}$ of $\pm 0.05$ are not significant. 
But, later while presenting our final results, we see that the ICAL detector can place competitive 
constraints on $\varepsilon_{\mu\tau}$ by exploiting the useful information contained in the spectral 
distributions of $\mu^-$ and $\mu^+$ events as a function of reconstructed observables $E_\mu$, $\cos\theta_\mu$, and $E'_{\rm had}$. 
To establish this claim, now, we show how the expected $\mu^-$ and $\mu^+$ event spectra  get modified in the 
presence of non-zero $\varepsilon_{\mu\tau}$ in terms of reconstructed $E_\mu$ and $\cos\theta_\mu$
while integrating over entire range of $E'_{\rm{had}}$. 
 \begin{figure}[h!]
 \begin{center}
 \begin{minipage}{.325\linewidth}
 \includegraphics[width=\textwidth]{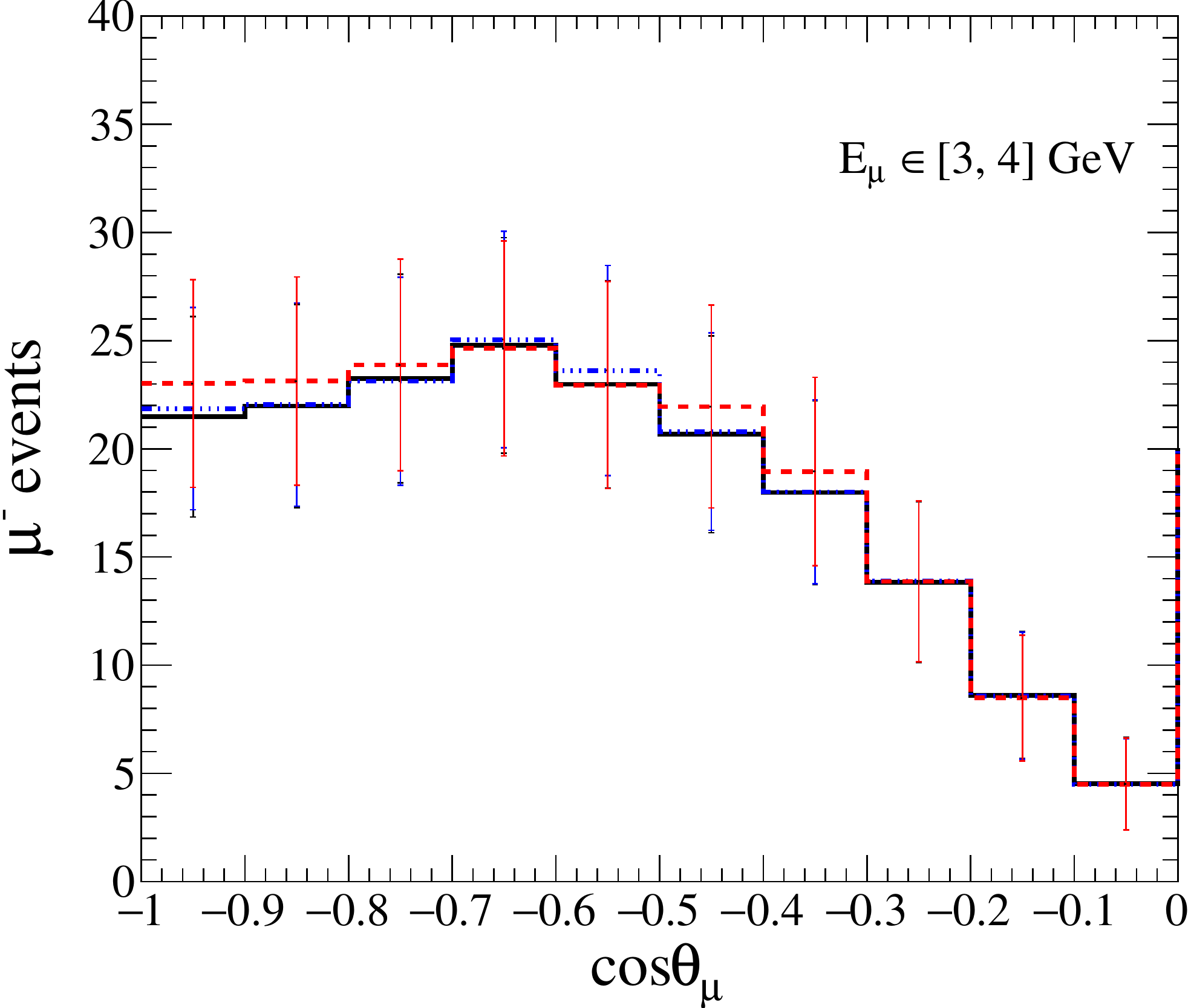} 
 \end{minipage}
 \hfill
 \begin{minipage}{.325\linewidth}
 \includegraphics[width=\textwidth]{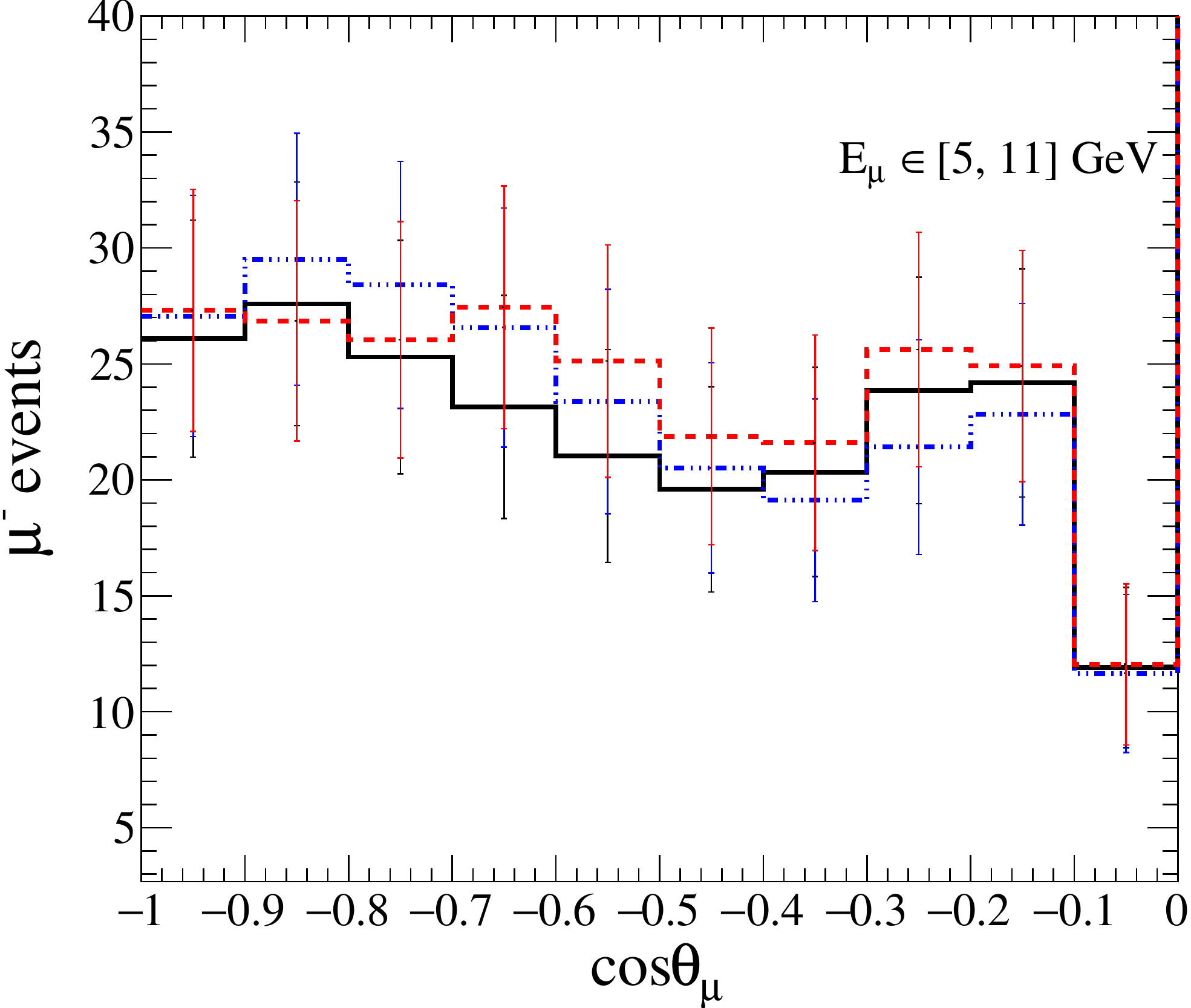} 
 \end{minipage}
 \hfill
  \begin{minipage}{.325\linewidth}
  \includegraphics[width=\textwidth]{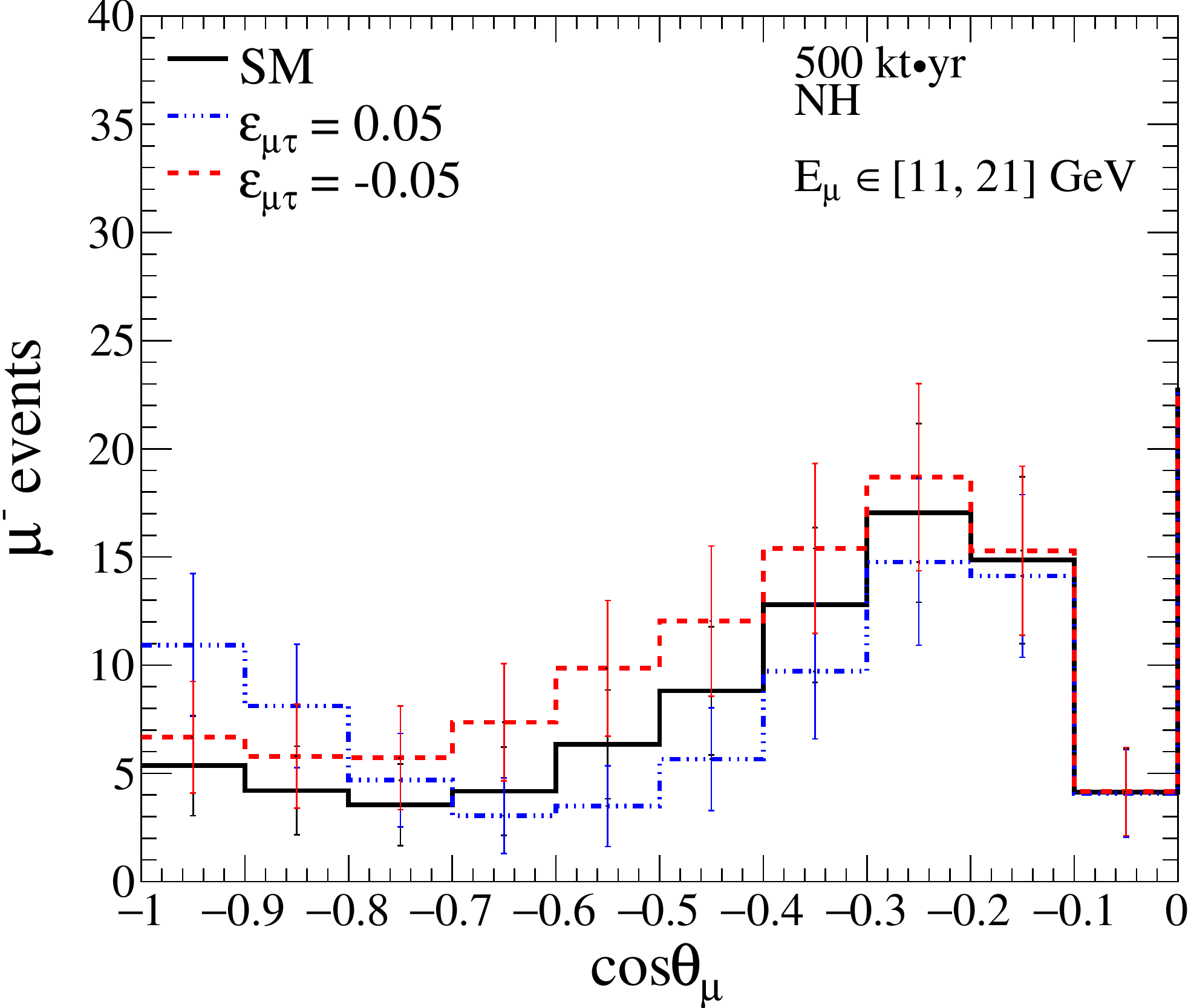} 
 \end{minipage}
 \hfill
 \begin{minipage}{.325\linewidth}
 \includegraphics[width=\textwidth]{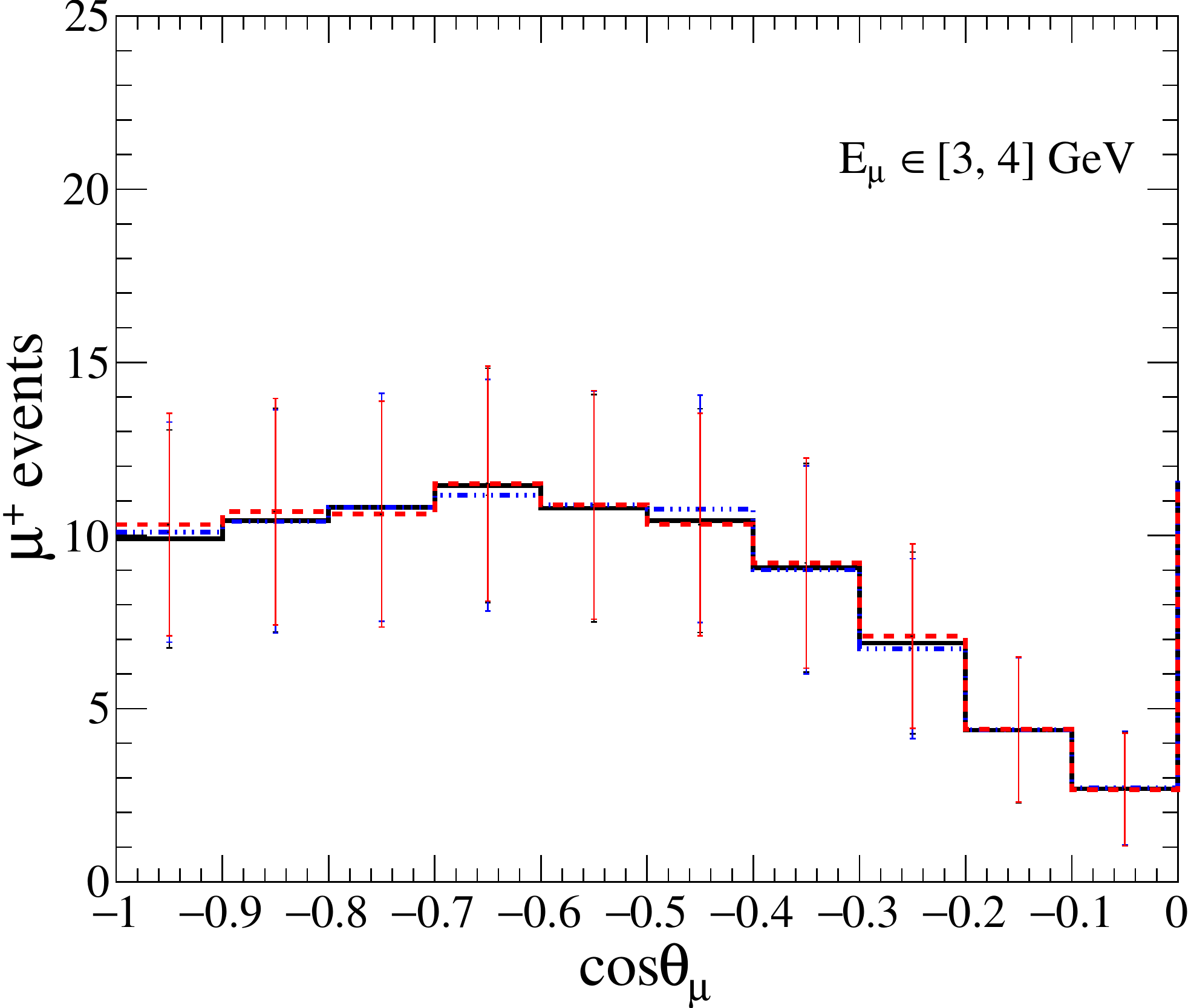} 
 \end{minipage}
\hfill
 \begin{minipage}{.325\linewidth}
 \includegraphics[width=\textwidth]{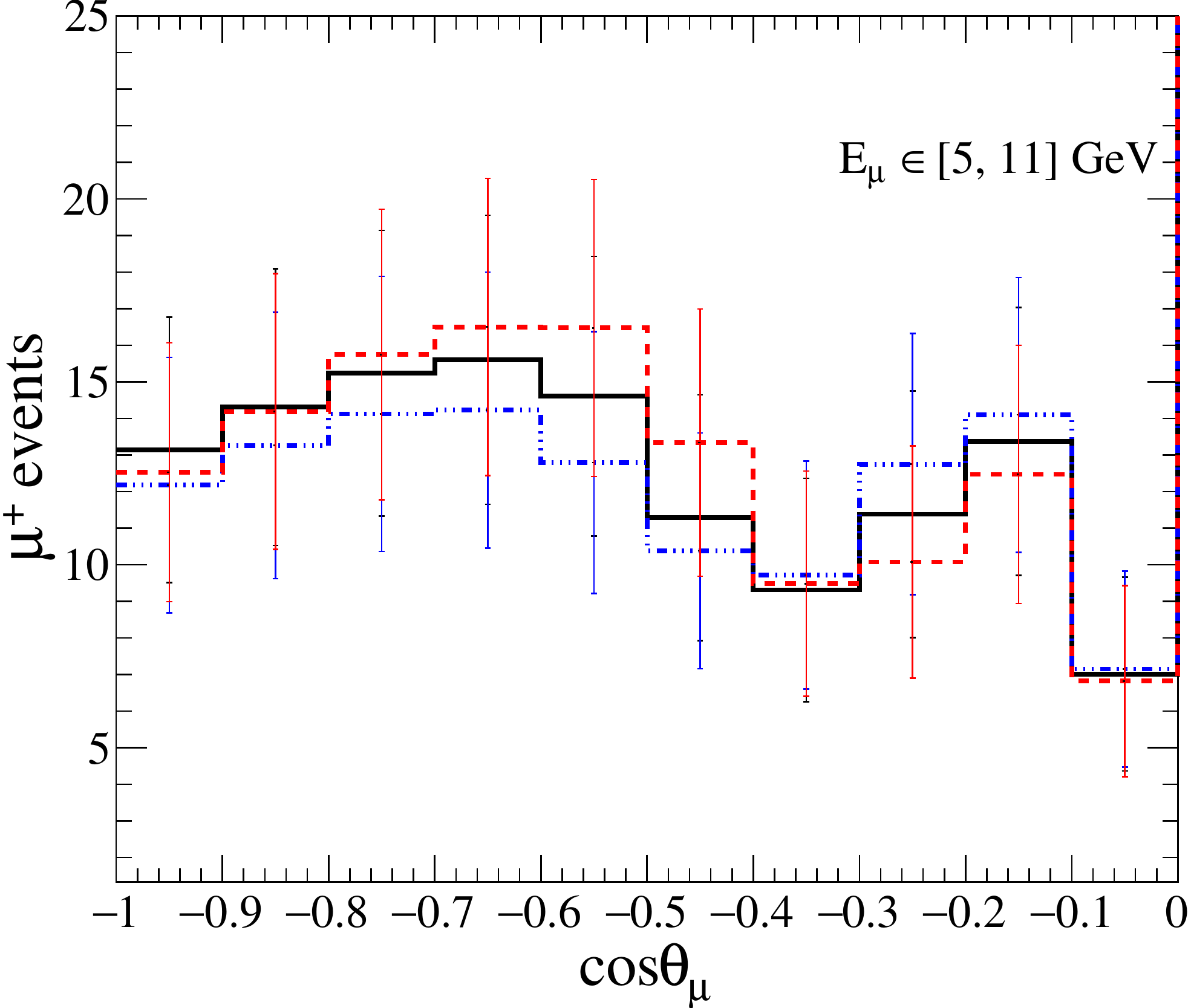} 
 \end{minipage}
 \hfill
  \begin{minipage}{.325\linewidth}
  \includegraphics[width=\textwidth]{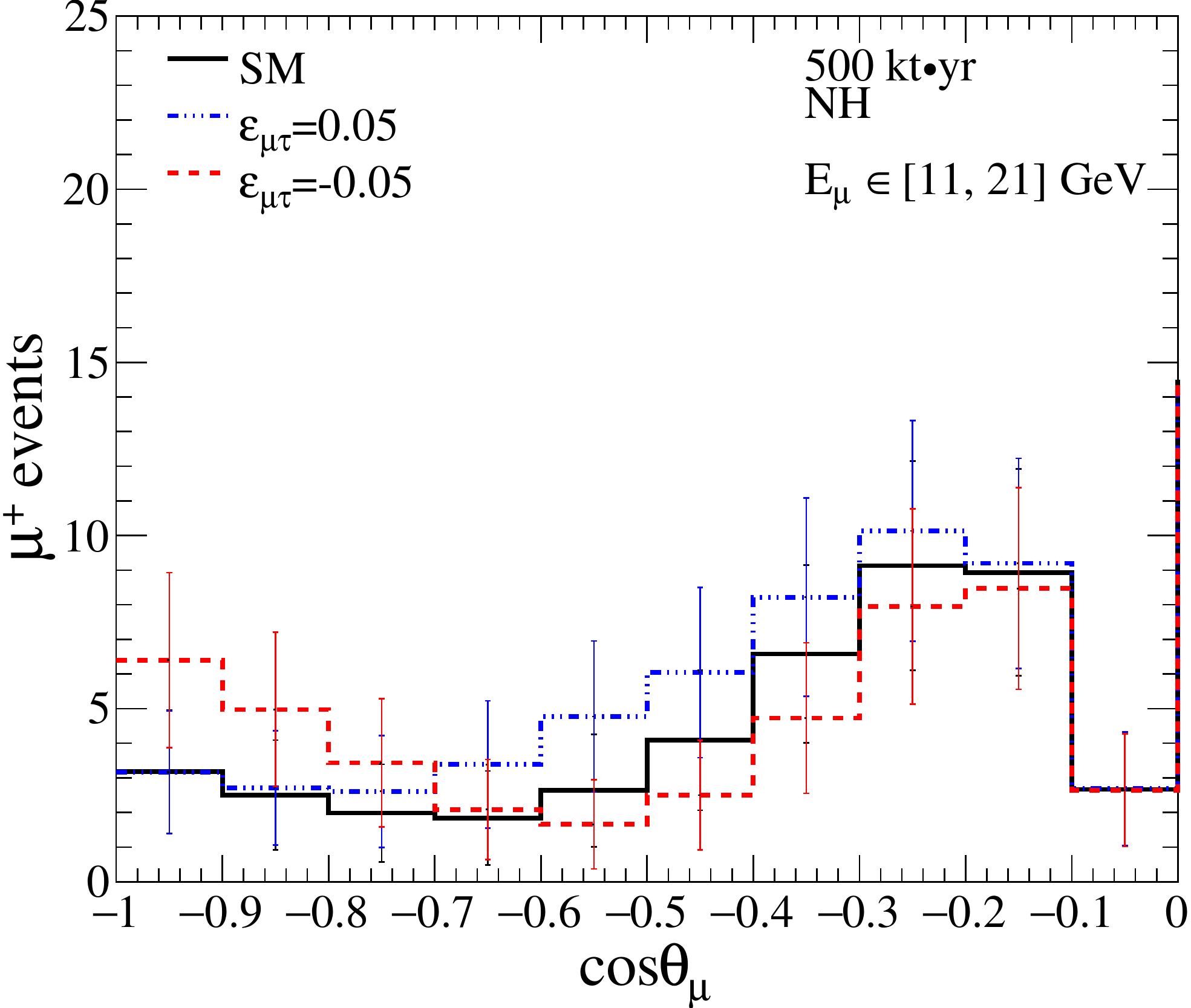} 
 \end{minipage}
 \mycaption{ The distributions of $\mu^-$ (upper panels) and $\mu^+$ (lower panels) events 
  for three different $E_\mu$ ranges: [3, 4]\,GeV in left panel, [5, 11]\,GeV in middle panel, 
  and [11, 21]\,GeV in right panel. In each panel, we consider three different cases: i) 
  $\varepsilon_{\mu\tau} = 0.05$ (blue line), ii) $\varepsilon_{\mu\tau} =0$ (the SM case, 
  black line), and iii) $\varepsilon_{\mu\tau} = -0.05$ (red line). Here, we sum over $E'_{\rm had}$ in its 
entire range of 0 to 25 GeV and show the results for 500 kt$\cdot$yr exposure and assuming NH. }
\label{fig:event-dist}
\end{center} 
\end{figure}
\subsection{Event Spectra }
In Fig.\,\ref{fig:event-dist}, we show the distributions of $\mu^-$ (upper panels) and $\mu^+$ 
(lower panels) events as a function of reconstructed $\cos\theta_\mu$ for three different ranges of $E_\mu$. The ranges of $E_\mu$ that we consider in left, middle, and right panels are [3, 4]\,GeV,\,\,[5, 11]\,GeV, 
and [11, 21]\,GeV respectively. Here, we integrate over $E'_{\rm{had}}$ in its entire range of 0 to 25\,GeV. 
In each panel, we compare the event spectra for three different cases:
i) $\varepsilon_{\mu\tau} = 0.05$ (blue line), ii)  $\varepsilon_{\mu\tau} =0$ (the SM case, black line), 
and iii) $\varepsilon_{\mu\tau} = -0.05$ (red line). 
Before we discuss the impact of non-zero $\varepsilon_{\mu\tau}$, we would like to mention few general 
features that are emerging from various panels in Fig.\,\ref{fig:event-dist}. For both $\mu^-$ (upper 
panels) and $\mu^+$ (lower panels), the number of events get reduced as we go to higher energies. It 
happens because of $\sim E_\nu^{-2.7}$ dependence of the atmospheric neutrino flux. Though the neutrino 
fluxes are higher along the horizontal direction ($\cos\theta_\mu$ around $0$) as compared to the other 
directions (for detailed discussion, see  Ref.\,\cite{Honda:2015fha}), but, due to the poor reconstruction 
efficiency of the ICAL detector along the horizontal direction\,\cite{Chatterjee:2014vta}, we see a suppression 
in $\mu^-$ and $\mu^+$ event rates around $\cos\theta_\mu\in[-0.1,\,0]$ irrespective of the choices of 
$E_\mu$ ranges. Important to note that as we proceed towards higher  $E_\mu$, the relative differences 
in $\mu^-$ and $\mu^+$ event rates   between the SM case ($\varepsilon_{\mu\tau} = 0$) and non-zero 
$\varepsilon_{\mu\tau}$ ($\pm 0.05$) get increased for a wide range of $\cos\theta_\mu$. We see similar 
features in Fig.\,\ref{fig:osc-dif} in Sec.\,\ref{sec:osc-prob-nsichap}, where we show the differences 
in $\nu_\mu\rightarrow\nu_\mu$ oscillograms due to the SM case ($\varepsilon_{\mu\tau} = 0$) and non-zero 
$\varepsilon_{\mu\tau}$ ($\pm 0.05$). We show the improvement in the sensitivity to constrain  
$\varepsilon_{\mu\tau}$ due to high energy events in Sec.\,\ref{sec:bounds-results-nsichap}. 
Next, we discuss the numerical technique and analysis procedure which we adopt to obtain the final results. 
\vfill
\section{Simulation Method}
\label{sec:method-analysis-nsi}
\subsection{Binning Scheme in ($E_{\mu}$, $\cos\theta_{\mu}$, $E'_{\rm had}$) Plane }
\label{sec:chap-nsi-binning}
In the present study, we produce all the results with low-energy (LE) and high-energy (HE) binning schemes. 
Table\,\ref{tab:bin-nsichap1} shows the detailed information about the LE binning scheme for the three 
reconstructed observables $E_{\mu}$, $\cos\theta_{\mu}$, and $E'_{\rm had}$. Table\,\ref{tab:bin-nsichap2} 
portrays the same for the HE binning scheme. 
In case of LE binning scheme, the range of $E_{\mu}$ is [1,\,11] GeV with total 10 bins each having a width of 1\,GeV.
In case of HE binning scheme, we extend the range of $E_{\mu}$ up to 21\,GeV by adding two additional bins in the range 
of 11 to 21\,GeV, where each bin has a width of 5 GeV. As far as reconstructed $E'_{\rm had}$ is concerned, in case of 
LE (HE) binning scheme, the considered range is 0 to 15 GeV (0 to 25 GeV). We can see from Table\,\ref{tab:bin-nsichap1} 
and Table\,\ref{tab:bin-nsichap2}  that the first three bins for $E'_{\rm had}$ are same for both the binning schemes,  
whereas the last bin extend from 4 to 15 GeV (4 to 25 GeV) for LE (HE) binning scheme. For both these binning schemes, we 
consider the entire range of $\cos\theta_{\mu}$ from -1 to 1. For upward going events, that is $\cos\theta_{\mu}\in\,$ [-1, 0], 
we consider 10 uniform bins each having width of 0.1. For downward going events,  that is $\cos\theta_{\mu}\in\,$ [0, 1],
we consider 5 uniform bins each having width of 0.2.  Important to note that the downward going events do not have 
enough path length to oscillate, but, these events play an important role to increase the overall statistics and to 
minimize the effect of normalization uncertainties in atmospheric neutrino fluxes. Here, we would like to mention that 
we have not optimized these binning schemes to obtain the best sensitivities, but we ensure that there are sufficient 
statistics in most of the bins. 

\begin{table}[htb!] 
\begin{center}
\begin{tabular}{|c| c| c| c| c|} 
\hline\hline 
Observable & Range & Bin width & No. of bins & Total bins \\ 
\hline
$E_{\mu}$ (GeV) & \makecell[c]{ $[1,11]$ } & \makecell[c]{ 1 } & \makecell{ 10 } & \makecell{10 } \\
\hline
$\cos\theta_\mu$ \,\, & \makecell[c]{ $[-1.0,0.0]$ \\ $[0.0,1.0]$ } & \makecell[c]{ 0.1 \\ 0.2} & 
\makecell[c]{10 \\ 5} & \makecell[c]{15}\\
\hline
$E'_{\rm had}$ (GeV) & \makecell[c]{$[0,2]$ \\ $[2,4]$ \\ $[4,15]$} 
& \makecell[c]{1 \\ 2 \\11 } & \makecell[c]{2 \\ 1 \\ 1} & \makecell[c]{4 }\\
\hline\hline
\end{tabular}
\end{center}
\mycaption{The low-energy (LE) binning scheme adopted for the reconstructed observables $E_\mu$, $\cos\theta_\mu$, and $E'_{\rm had}$ 
for each muon polarity. The last column shows the total number of bins taken for each observable.}
\label{tab:bin-nsichap1}
\end{table}
\begin{table}[htb!] 
\begin{center}
\begin{tabular}{|c| c| c| c| c| } 
\hline\hline 
Observable & Range & Bin width & No. of bins & Total bins \\ 
\hline
$E_{\mu}$ (GeV) & \makecell[c]{ $[1,11]$ \\$[11,21]$} & 
\makecell[c]{ 1 \\ 5} & 
\makecell{ 10 \\2 } & \makecell{12 } 
 \\
\hline
$\cos\theta_\mu$\,\,  & \makecell[c]{ $[-1.0,0.0]$ \\ $[0.0,1.0]$ } 
& \makecell[c]{ 0.1 \\ 0.2} & 
\makecell[c]{10 \\ 5} & \makecell[c]{15 }\\
\hline
$E'_{\rm had}$ (GeV) & \makecell[c]{$[0,2]$ \\ $[2,4]$ \\ $[4,25]$} 
& \makecell[c]{1 \\ 2  \\ 21 } & 
\makecell[c]{2 \\ 1 \\ 1} & \makecell[c]{4 }\\
\hline\hline
\end{tabular}
\end{center}
\mycaption{The high-energy (HE) binning scheme considered for the reconstructed observables $E_\mu$, $\cos\theta_\mu$, 
and $E'_{\rm had}$ for each muon polarity. The last column shows the total number of bins taken for each observable.}
\label{tab:bin-nsichap2}
\end{table}
\subsection{Numerical Analysis}
\label{sec:numerical-nsichap}
In our analysis, the $\chi^2$ function gives us the median sensitivity of the experiment in the frequentist 
approach\,\cite{Blennow:2013oma}. We use the following  Poissonian $\chi^2_{-}$ for $\mu^-$ events 
in our statistical analysis considering $E_\mu$, $\cos\theta_\mu$, and $E'_{\rm had}$ as observables (the  
so-called ``3D'' analysis as considered in\,\cite{Devi:2014yaa}):  
\begin{equation}
 \chi^2_{-}({\rm 3D})\,=\,\min_{\zeta_l}\,\sum^{N_{E'_{\rm had}}}_{i=1} 
 \sum^{N_{E_\mu}}_{j=1} \sum^{N_{{\cos\theta_\mu}}}_{k=1}
 \,2\,\bigg[ N^{\rm theory}_{ijk}\,-
 \, N^{\rm data}_{ijk}\, - \,  N^{\rm data}_{ijk}\,\,
 {\rm{ln}}\,\bigg(\frac{N^{\rm theory}_{ijk}}{N^{\rm data}_{ijk}}
 \bigg)\bigg]\,+\,\sum^5_{l=1} \zeta^2_{l}\,,
\label{eq:chisq-chapnsi}
\end{equation}
with 
\begin{equation}
 N^{\rm theory}_{ijk}\,=\,N^0_{ijk}\big(\,1\,+\,\sum^5_{l=1}
 \pi^l_{ijk}\zeta_{l}\,\big)\,.
 \label{sec:chisqICAL-chapnsi}
\end{equation}
In the above equations, $N^{\rm data}_{ijk}$ and $N^{\rm theory}_{ijk}$ denote the observed and 
expected number of $\mu^-$ events in a given [$E_\mu$, $\cos\theta_\mu$, $E'_{\rm had}$] bin. In case of 
LE (HE) binning scheme, $N_{E_\mu}$ = 10 (12), $N_{\cos\theta_\mu} = 15$, and $N_{E'_{\rm had}} = 4$. In 
Eq.\,\ref{sec:chisqICAL-chapnsi}, $N^0_{ijk}$ represents the number of expected events without systematic 
uncertainties. Following Ref.\,\cite{Ghosh:2012px}, we consider five systematic errors in our analysis: 20$\%$ flux normalization error, 
10$\%$ error in cross-section, 5$\%$ tilt error, 5$\%$ zenith angle error, and 5$\%$ overall systematics. 
We incorporate these systematic uncertainties in our simulation using the well known ``pull'' 
method\,\cite{Huber:2002mx,Fogli:2002pt,GonzalezGarcia:2004wg}. In Eq.\,\ref{eq:chisq-chapnsi} and 
Eq.\,\ref{sec:chisqICAL-chapnsi}, the quantities $\zeta_{l}$ denote the ``pulls'' due to the systematic
uncertainties. 

When we produce results with only $E_\mu$ and  $\cos\theta_\mu$ as observables and do not use the 
information on hadron energy $E'_{\rm had}$ (the so-called ``2D'' analysis as considered in 
Ref.\,\cite{Ghosh:2012px}), the Poissonian $\chi^2_{-}$ for $\mu^-$ events takes the form 
\begin{equation}
 \chi^2_{-}({\rm 2D})\,=\,\min_{\zeta_l}\,\sum^{N_{E_\mu}}_{j=1} \sum^{N_{{\cos\theta_\mu}}}_{k=1}
 \,2\,\bigg[ N^{\rm theory}_{jk}\,-
 \, N^{\rm data}_{jk}\, - \,  N^{\rm data}_{jk}\,\,
 {\rm{ln}}\,\bigg(\frac{N^{\rm theory}_{jk}}{N^{\rm data}_{jk}}
 \bigg)\bigg]\,+\,\sum^5_{l=1} \zeta^2_{l}\,,
\label{eq:chisq-chapnsi-2d}
\end{equation}
with 
\begin{equation}
 N^{\rm theory}_{jk}\,=\,N^0_{jk}\big(\,1\,+\,\sum^5_{l=1}
 \pi^l_{jk}\zeta_{l}\,\big)\,.
 \label{sec:chisqICAL-chapnsi-2d}
\end{equation}
In Eq.\,\ref{eq:chisq-chapnsi-2d}, $N^{\rm data}_{jk}$ and $N^{\rm theory}_{jk}$ indicate the observed and 
expected number of $\mu^-$ events in a given [$E_\mu$, $\cos\theta_\mu$] bin. In Eq.\,\ref{sec:chisqICAL-chapnsi-2d}, 
$N^0_{jk}$ stands for the number of expected events without systematic errors. In case of 
LE (HE) binning scheme, $N_{E_\mu}$ = 10 (12) and $N_{\cos\theta_\mu} = 15$.

For both the ``2D'' and ``3D''  analyses, the  $\chi^2_+$ for $\mu^+$ events is determined following 
the same technique described above.  We add the individual contributions from $\mu^-$ and $\mu^+$ 
events to estimate the total $\chi^2$ for both the ``2D'' and ``3D'' schemes:   
\begin{equation}
\chi^2_{\rm ICAL} = \chi^2_-\,+\,\chi^2_+\,.
\label{eq:total-chisq}
\end{equation}

In our analysis, we simulate the prospective data
considering the following benchmark values of the oscillation parameters: $\sin^2 \theta_{23} = 0.5$, 
$\,\sin^2 2\theta_{13}= 0.1$,  $\sin^2 \theta_{12}\, =\, 0.3$, $\Delta m^2_{21}\,
=\, 7.5\times10^{-5}\,\rm{eV^2}$, and $|\Delta m^2_{\rm eff}|\, =\, 2.4\times10^{-3}\, 
\rm{eV^2}$. To estimate the value of $\Delta m^2_{31}$ 
from $\Delta m_{\rm eff}^{2}$, we use the Eq.\,\ref{eq:dmseff-chap3}, where $\Delta 
m_{\textrm{eff}}^{2}$ has the same magnitude for NH and IH with +ve and -ve signs 
respectively. 
In the fit, we first minimize $\chi^2_{\rm ICAL}$ (see Eq.\,\ref{eq:total-chisq}) 
with respect to the ``pull'' variables $\zeta_l$,  
and then marginalize over the oscillation parameters  $\sin^2\theta_{23}$ 
in the range [0.36,\,0.66], $|\Delta m_{\textrm{eff}}^{2}|$  in the range 
$[2.1, 2.6]\times 10^{-3}$ eV$^2$, and over both the choices of mass hierarchy, NH and IH, while keeping 
$\theta_{12}$, $\Delta m^2_{21}$, $\sin^2 2\theta_{13}$ fixed at their 
benchmark values. We consider $\delta_{\rm CP} = 0^{\circ}$ throughout our analysis.

\section{Results}
\label{sec:results-nsichap}

\subsection{Expected Bounds on NSI parameter $\varepsilon_{\mu\tau}$}
\label{sec:bounds-results-nsichap}

We quantify the statistical significance of the analysis to constrain 
the NSI parameter $\varepsilon_{\mu\tau}$ in the following fashion
\begin{equation}
 \Delta \chi^2_{\,\,{\rm ICAL-NSI}} = \chi^2_{\,\,{\rm ICAL}}\left({\rm SM} + \varepsilon_{\mu\tau}\right)
 -\chi^2_{\,\,{\rm ICAL}}\left({\rm SM}\right)\,.  
\end{equation}
Here, $\chi^2_{\,\,{\rm ICAL}} ({\rm SM})$ and $\chi^2_{\,\,{\rm ICAL}}\left({\rm SM} + 
\varepsilon_{\mu\tau}\right)$ are calculated by fitting the prospective data   
with zero (the SM case) and non-zero value of NSI parameter $\varepsilon_{\mu\tau}$ 
respectively. In our analysis procedure, statistical fluctuations are suppressed, and therefore, 
$\chi^2_{\,\,{\rm ICAL}}({\rm SM})\approx 0$.  

 \begin{figure}[htb!]
  \begin{center}
 \begin{minipage}{.48\linewidth}
 \includegraphics[width=\textwidth]{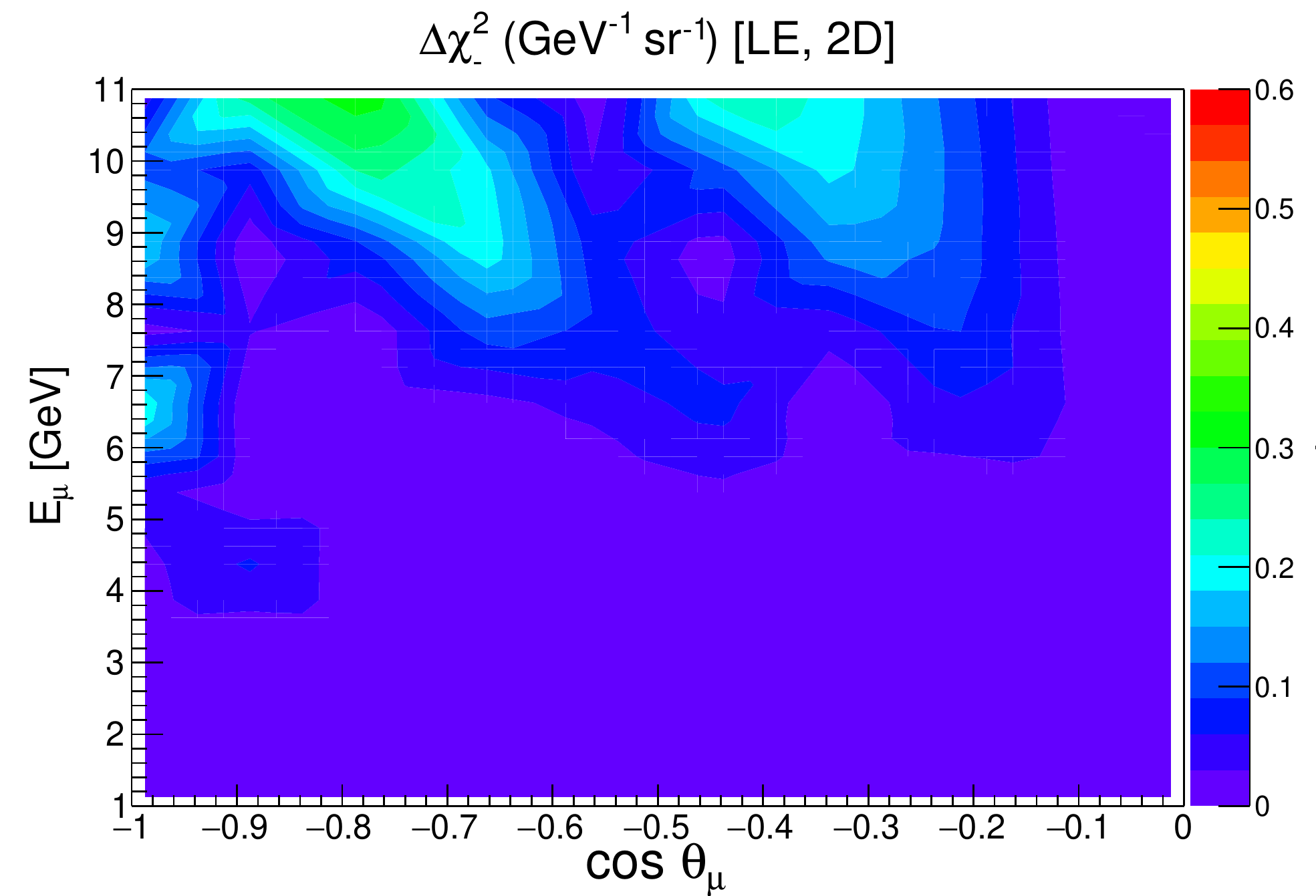} 
 \end{minipage}
 \hfill
 \begin{minipage}{.48\linewidth}
 \includegraphics[width=\textwidth]{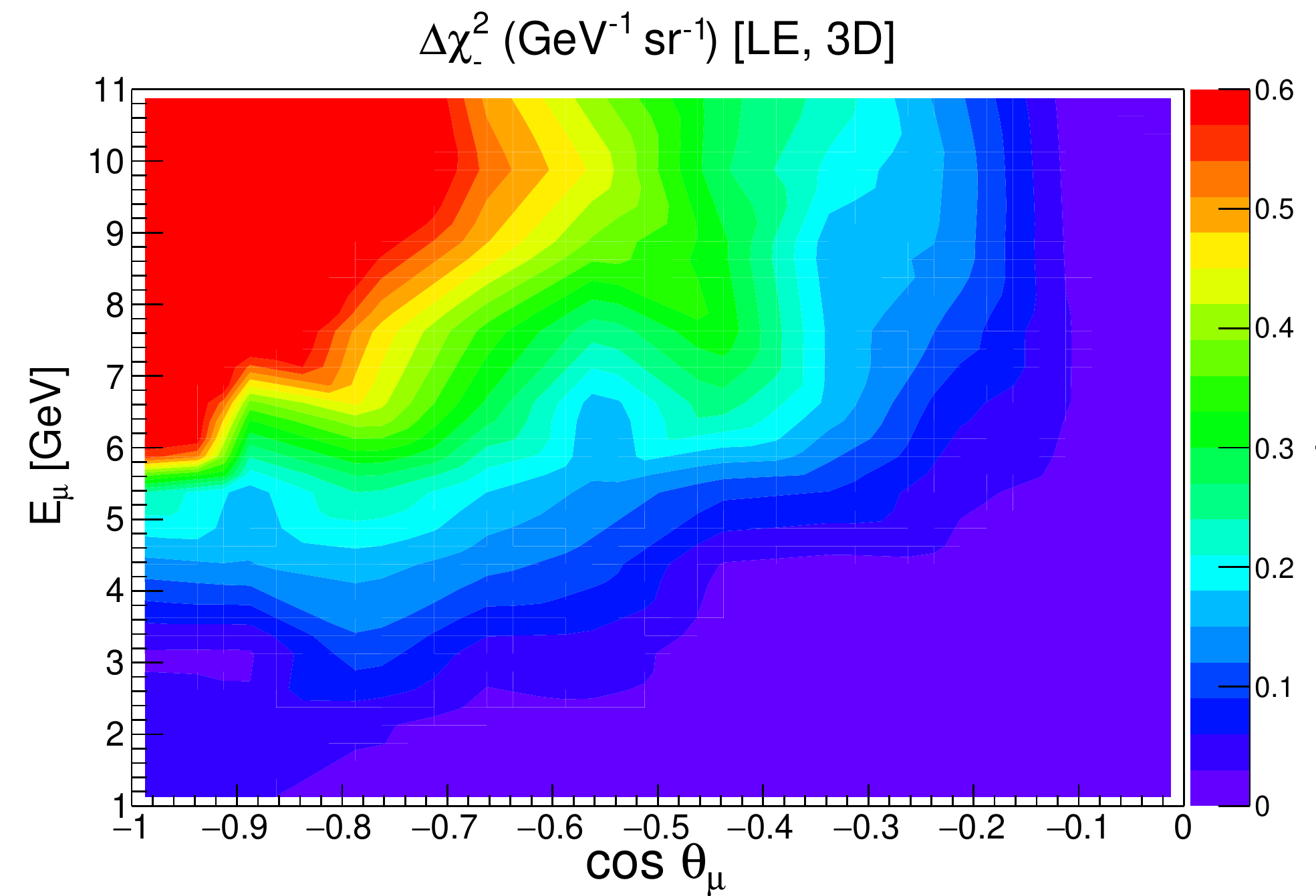} 
 \end{minipage}
 \hfill
 \begin{minipage}{.48\linewidth}
 \includegraphics[width=\textwidth]{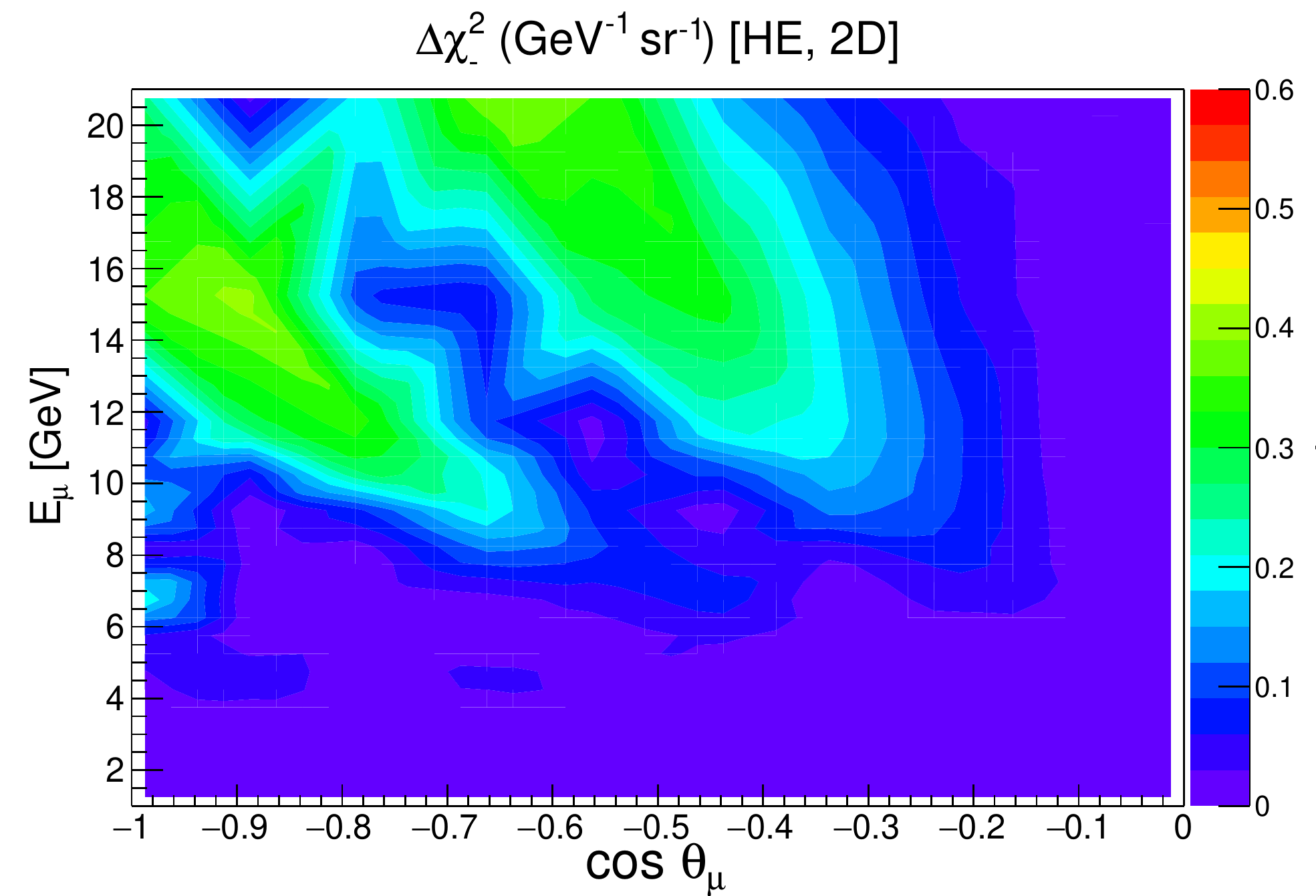} 
 \end{minipage}
 \hfill
 \begin{minipage}{.48\linewidth}
 \includegraphics[width=\textwidth]{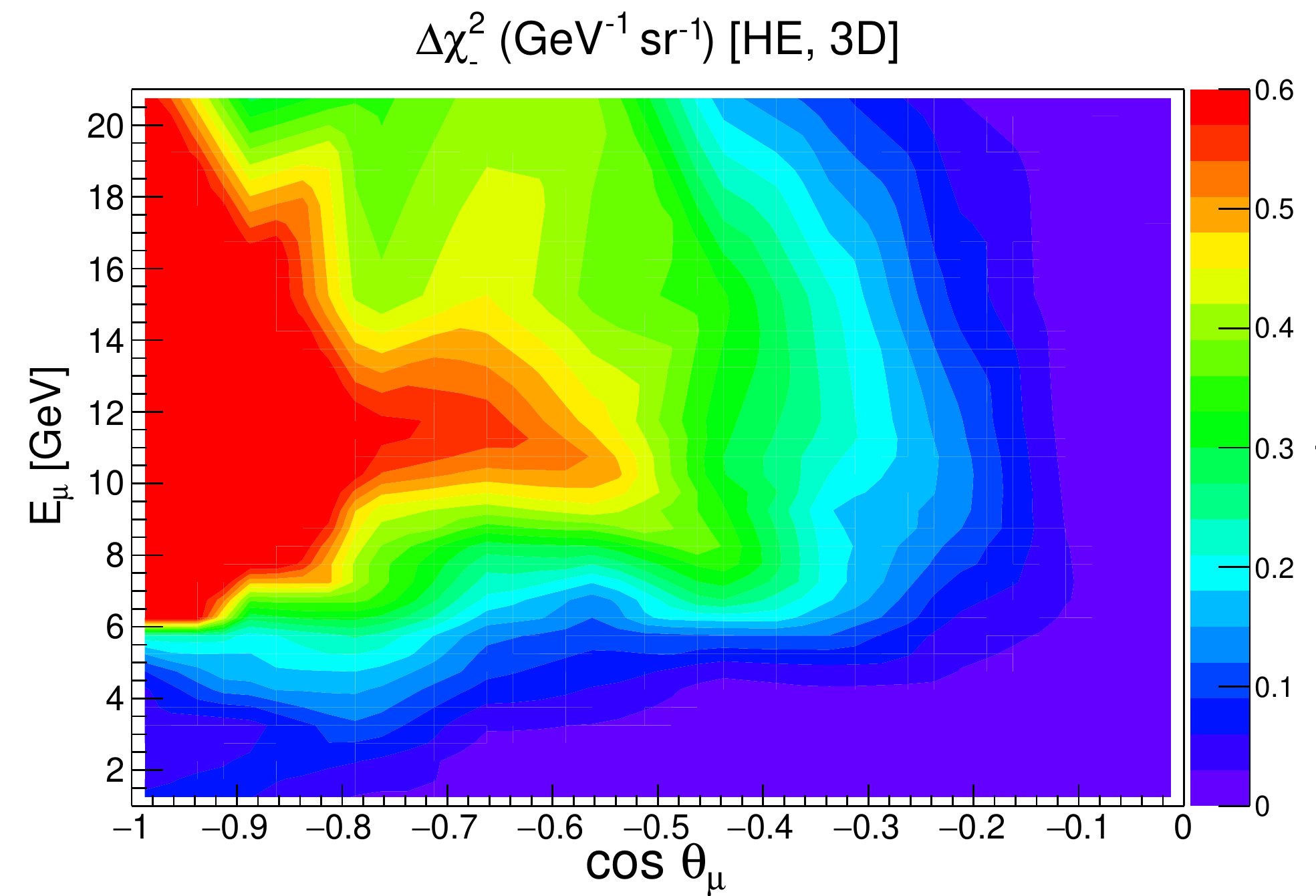} 
 \end{minipage}
 \mycaption{ Distributions of $\Delta\chi^2_{\rm ICAL-NSI}$ (per unit area) from $\mu^-$ events in reconstructed $\cos\theta_\mu$ 
 and $E_\mu$ plane assuming non-zero $\varepsilon_{\mu\tau}$ in the fit with a strength of $0.05$. The top 
 (bottom) panels are for the LE (HE) binning scheme. For a given binning scheme, left and right panels are 
 obtained with  [$E_\mu$,  $\cos\theta_\mu$] and [$E_\mu$, $\cos\theta_\mu$, $E'_{\rm had}$] respectively.   
In all the panels, we use 500 kt$\cdot$yr exposure and assume NH in both data and theory. }
\label{fig:E-vs-costhe-nu}
\end{center} 
\end{figure}
\begin{figure}[htb!]
  \begin{center} 
 \begin{minipage}{.48\linewidth}
  \includegraphics[width=\textwidth]{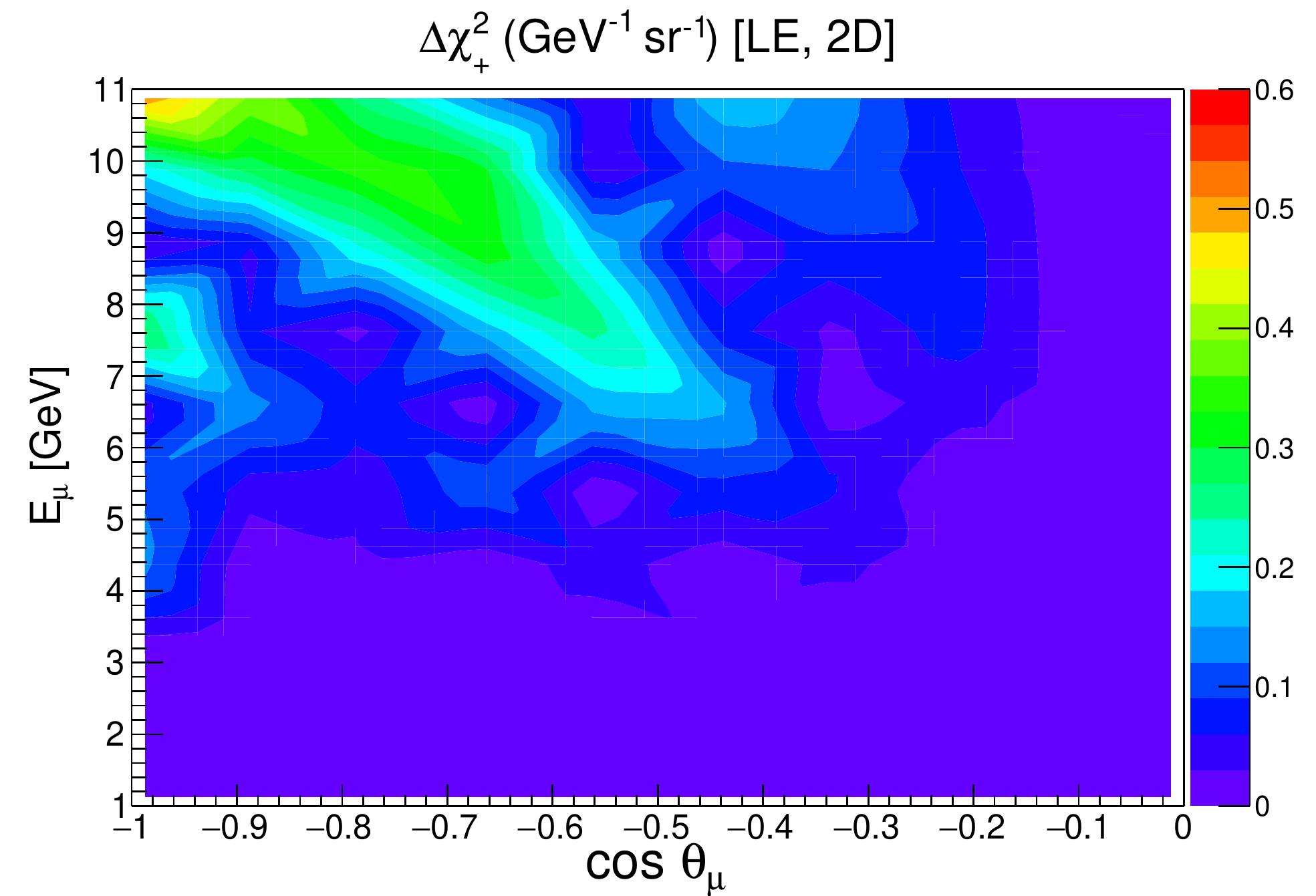} 
 \end{minipage}
 \hfill
 \begin{minipage}{.48\linewidth}
 \includegraphics[width=\textwidth]{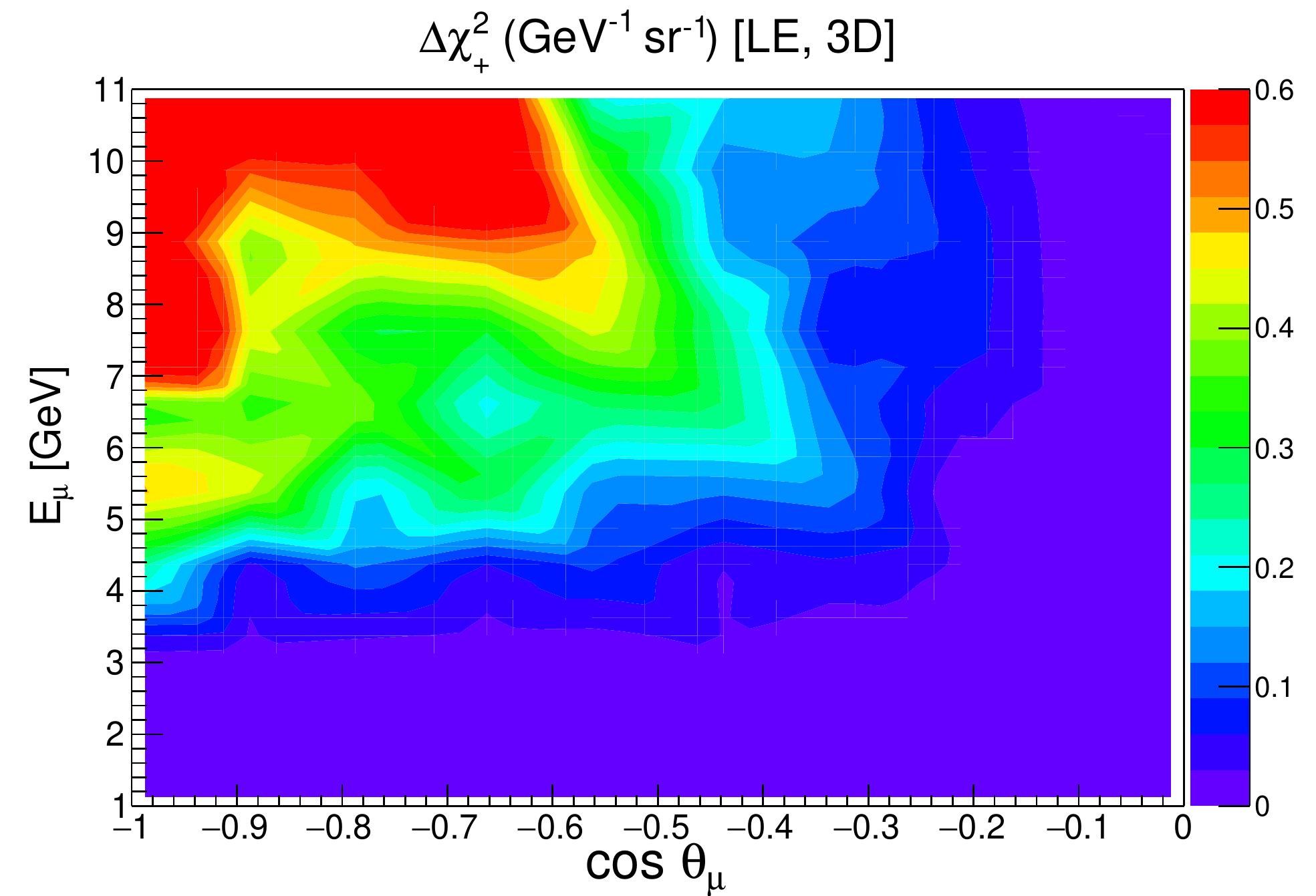} 
 \end{minipage}
 \hfill
  \begin{minipage}{.48\linewidth}
  \includegraphics[width=\textwidth]{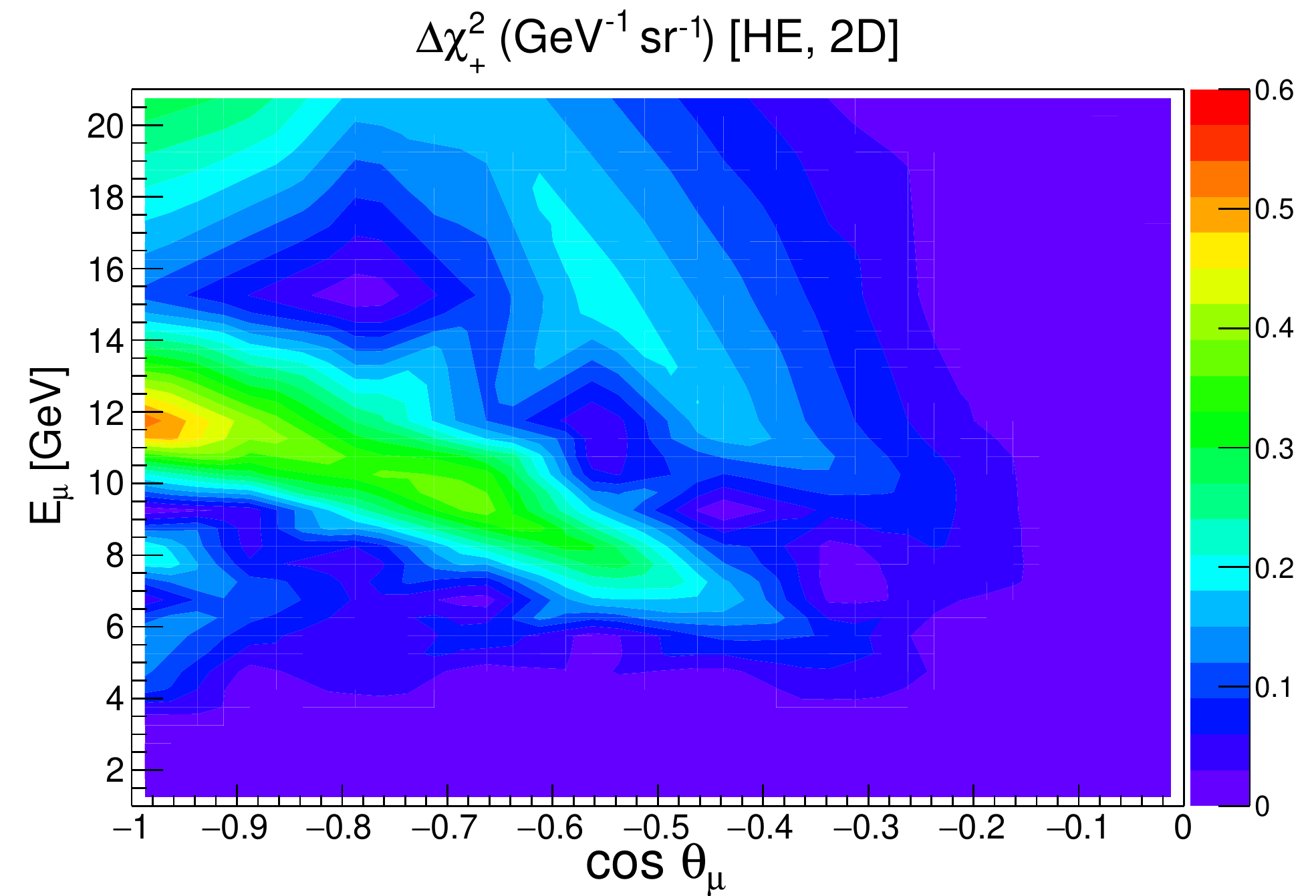} 
 \end{minipage}
 \hfill
 \begin{minipage}{.48\linewidth}
 \includegraphics[width=\textwidth]{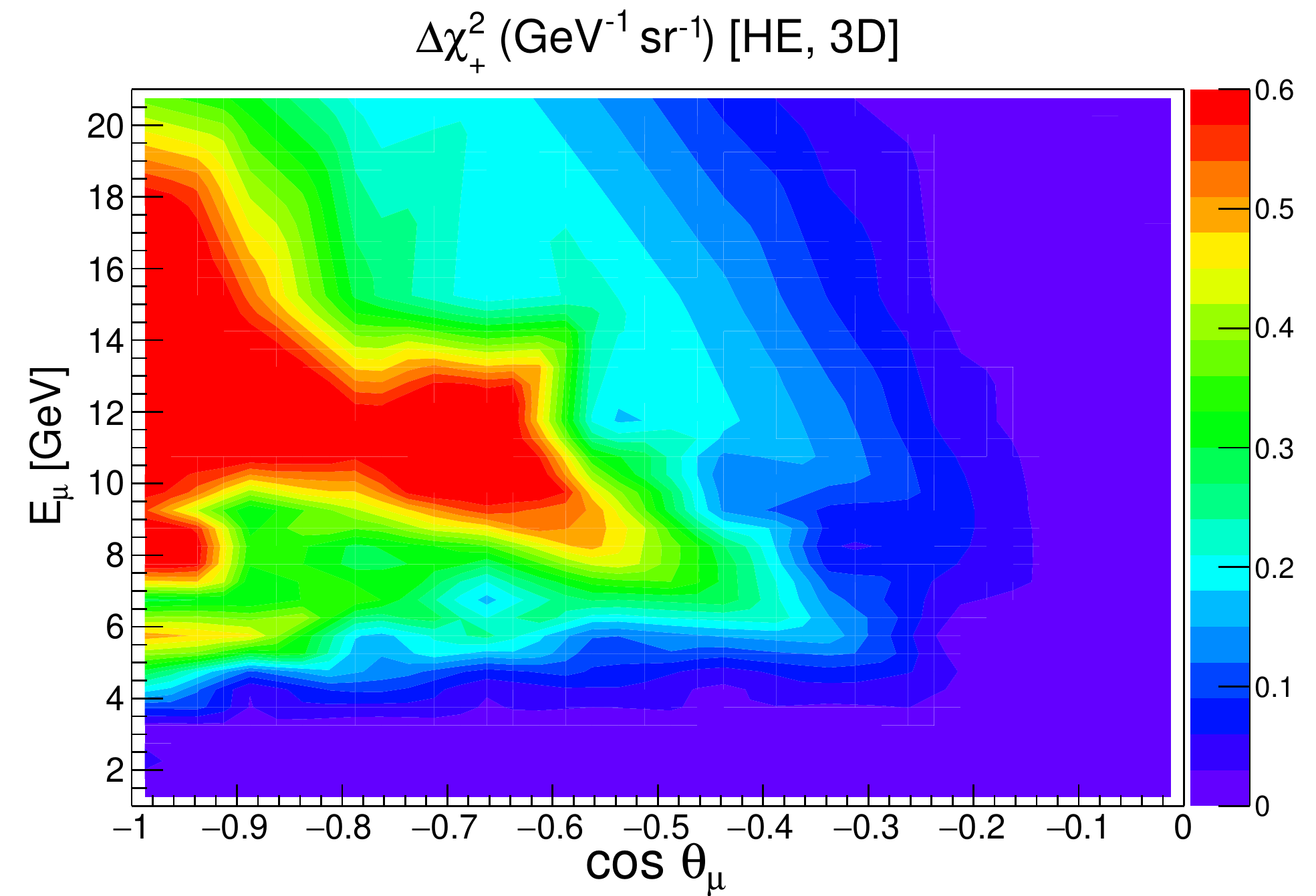} 
 \end{minipage}
 \mycaption{ Distributions of $\Delta\chi^2_{\rm ICAL-NSI}$ (per unit area) from $\mu^+$ events in reconstructed $\cos\theta_\mu$ 
 and $E_\mu$ plane assuming non-zero $\varepsilon_{\mu\tau}$ in the fit with a strength of $0.05$. The top 
 (bottom) panels are for the LE (HE) binning scheme. For a given binning scheme, left and right panels are 
 obtained with  [$E_\mu$,  $\cos\theta_\mu$] and [$E_\mu$, $\cos\theta_\mu$, $E'_{\rm had}$] respectively. 
In all the panels, we use 500 kt$\cdot$yr exposure and assume NH in both data and theory. }
\label{fig:E-vs-costhe-anu}
\end{center} 
\end{figure}

Let us first identify the regions in $\cos\theta_\mu$ and $E_\mu$ plane which give significant contributions
towards $\Delta\chi^2_{\rm ICAL-NSI}$. In Fig.\,\ref{fig:E-vs-costhe-nu}, we show the distribution\footnote{In 
Fig.\,\ref{fig:E-vs-costhe-nu}, we do not consider the constant contributions in $\chi^2$ coming from the term 
which involves five pull parameters $\zeta^2_l$ in Eq.\,\ref{eq:chisq-chapnsi} and Eq.\,\ref{eq:chisq-chapnsi-2d}.
Also, we do not marginalize over the oscillation parameters in the fit to produce these figures. We adopt the same 
strategy for Fig.\,\ref{fig:E-vs-costhe-anu} as well. Note that we show our 
final results considering full pull contributions and marginalizing over the oscillation parameters in the 
fit as mentioned in previous section. } of $\Delta \chi^2_{-}$ from $\mu^-$ events in the reconstructed [$\cos\theta_\mu$-$E_\mu$]
plane using 500 kt$\cdot$yr exposure of the ICAL detector and assuming NH. In all the panels of 
Fig.\,\ref{fig:E-vs-costhe-nu}, we consider $\varepsilon_{\mu\tau} = 0.05$ in the fit and show the results 
for the following four different choices of binning schemes and observables: i) top left panel: [LE, 2D], ii) 
top right panel: [LE, 3D], iii) bottom left panel: [HE, 2D], iv) bottom right panel: [HE, 3D]. We show the distribution 
of $\Delta \chi^2_{+}$ from $\mu^+$ events in the plane of reconstructed $\cos\theta_\mu$ and $E_\mu$
for these four different cases in  Fig.\,\ref{fig:E-vs-costhe-anu} considering $\varepsilon_{\mu\tau} = 0.05$ in the fit. 
In left panels of Figs.\,\ref{fig:E-vs-costhe-nu} and 
Fig.\,\ref{fig:E-vs-costhe-anu}, we show the sensitivity in the plane of 
reconstructed $\cos\theta_\mu$ and $E_\mu$ for the ``2D'' analysis, where we do not use any information on hadrons. 
But, in right panels of these figures, we portray the sensitivity in the plane of 
reconstructed $\cos\theta_\mu$ and $E_\mu$ for the ``3D'' case, where the events are further divided into four sub-bins 
depending on the reconstructed hadron energy for LE (see Table\,\ref{tab:bin-nsichap1}) and HE binning schemes 
(see Table\,\ref{tab:bin-nsichap2}). 

The common features which are emerging from all the panels in Fig.\,\ref{fig:E-vs-costhe-nu} and 
Fig.\,\ref{fig:E-vs-costhe-anu} are that most of the sensitivity towards the NSI parameter $\varepsilon_{\mu\tau}$ 
stems from higher energies and longer baselines where the matter effect term 2$\sqrt{2} G_F N_e E$ becomes 
sizeable. We observe similar trends in Fig\,\ref{fig:osc-dif} where we plot the differences in $\nu_\mu\rightarrow\nu_\mu$ 
oscillation probabilities for the cases $\varepsilon_{\mu\tau} = 0$ and $\varepsilon_{\mu\tau} = \pm 0.05$. 
The event spectra as shown in  Fig.\,\ref{fig:event-dist}  also confirm this fact. Fig.\,\ref{fig:E-vs-costhe-nu} and 
Fig.\,\ref{fig:E-vs-costhe-anu} clearly demonstrate while going from LE to HE binning scheme that the sensitivity towards 
the NSI parameter $\varepsilon_{\mu\tau}$ get enhanced due to the increment in the range of $E_\mu$ from 11 GeV to 21 GeV
and for extending the fourth $E'_{\rm had}$ bin from 15 GeV to 25 GeV. 
We can also observe from these figures that with the addition of  hadron energy information, the area in the $E_\mu$-$\cos\theta_\mu$ 
plane which contributes significantly to $\Delta \chi^2_{\pm}$ increases, consequently enhancing the net $\Delta \chi^2_{\pm}$
for both LE and HE binning schemes. Here, we would like to mention that the increase in  $\chi^2_{\pm}$ is not just 
due to the information contained in $E'_{\rm had}$, but also due to the valuable information coming from the correlation 
between $E'_{\rm had}$ and muon momentum ($E_\mu$, $\cos\theta_\mu$).   
 
 \begin{figure}[t!]
  \begin{center}
  \begin{minipage}{.50\linewidth}
  \includegraphics[width=\textwidth]{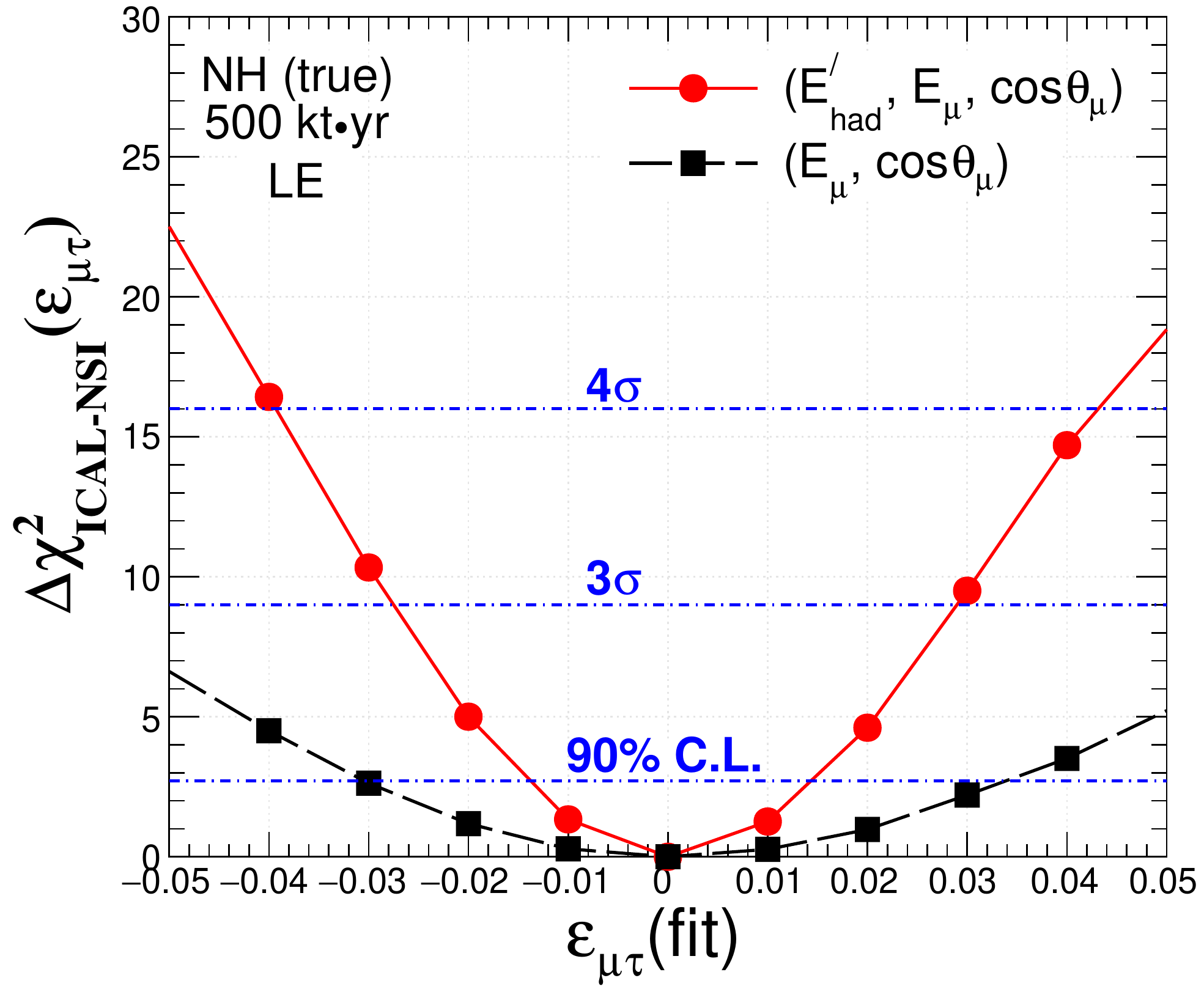}
  \end{minipage}
  \hfill
  \begin{minipage}{.49\linewidth}
   \includegraphics[width=\textwidth]{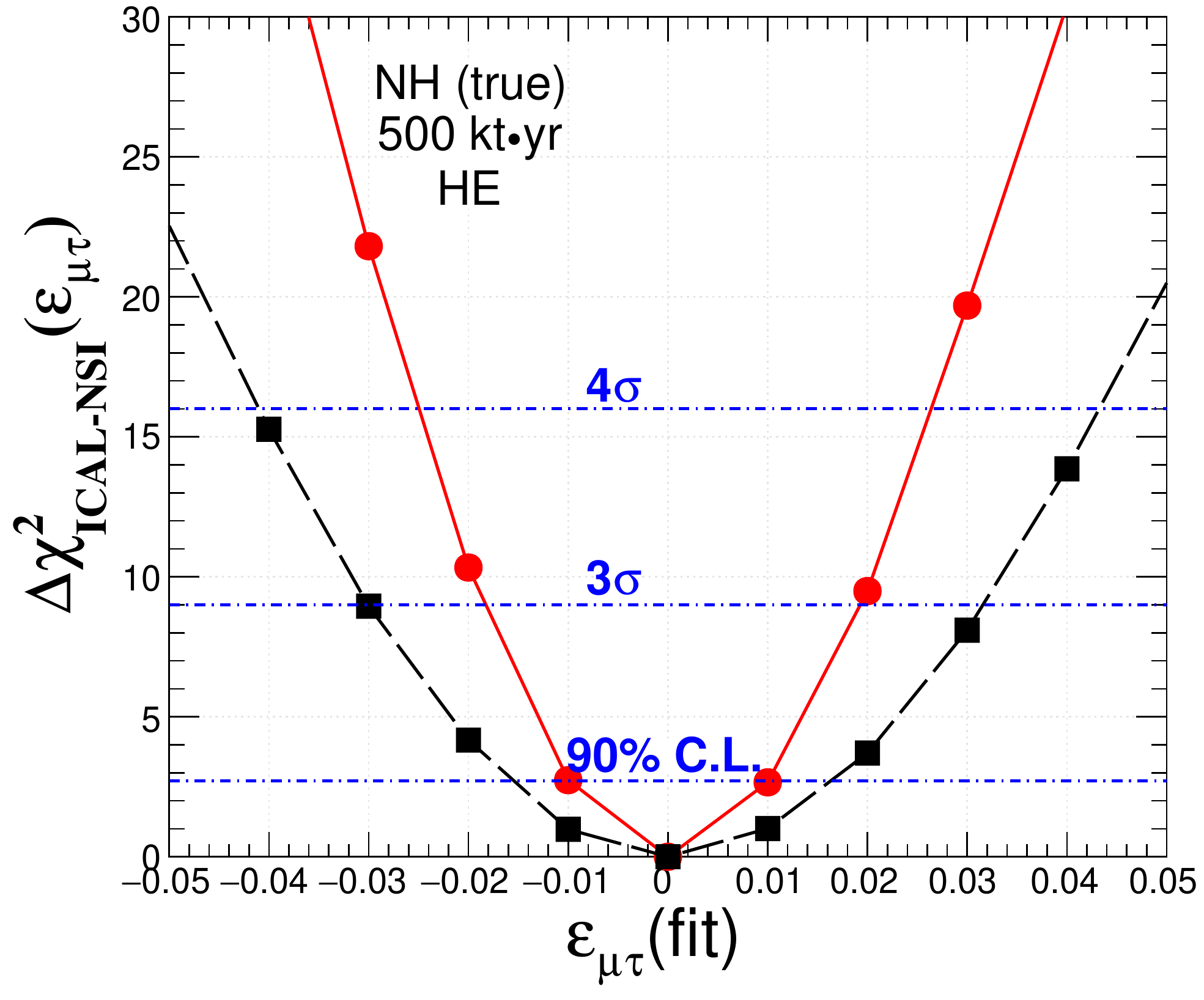}
  \end{minipage}
\mycaption{The sensitivity of the ICAL detector to set bounds on the NSI parameter 
$\varepsilon_{\mu\tau}$ using 500 kt$\cdot$yr exposure and assuming NH.  
Left (right) panel is with LE (HE) binning scheme. In each panel, the red solid line 
shows the sensitivity for the ``3D'' where we consider  $E_{\mu}$,\,$\cos\theta_{\mu}$, 
and $E'_{\rm had}$ as observables. The black dashed line in each panel portrays 
the sensitivity for the ``2D'' case considering $E_{\mu}$  and $\cos\theta_{\mu}$ 
as observables. These results are obtained after performing marginalization 
over $\theta_{23}$, $\Delta m^{2}_{\rm eff}$, and both choices of mass hierarchy. }
 \label{fig:bound-nh}
\end{center} 
\end{figure}

In Fig.\,\ref{fig:bound-nh}, we show the sensitivity of the ICAL detector to constrain 
$\varepsilon_{\mu\tau}$ using an exposure of 500 kt$\cdot$yr and assuming NH as the true 
mass hierarchy. We obtain these results after performing marginalization over $\theta_{23}$, 
$\Delta m^{2}_{\rm eff}$, and both the choices of mass hierarchy as discussed in 
Sec.\,\ref{sec:numerical-nsichap}. 
In the left (right) panel, the results are shown for the LE (HE) binning scheme. In each panel, 
the red solid  line shows the sensitivity for the ``3D'' case where we consider  $E_{\mu}$, $\cos\theta_{\mu}$, 
and $E'_{\rm had}$ as observables. The black dashed line in each panel portrays the sensitivity 
for the ``2D'' case considering $E_{\mu}$  and $\cos\theta_{\mu}$ as observables. We see considerable 
improvement in the sensitivity for both the LE and HE binning schemes when we add $E'_{\rm had}$ 
along with $E_\mu$ and $\cos\theta_\mu$ as observables. We see significant gain in the sensitivity when we 
increase the $E_\mu$ range from 11 GeV to 21 GeV and extend the fourth $E'_{\rm had}$ bin from 15 GeV to 
25 GeV. It is evident from both the panels in Fig.\,\ref{fig:bound-nh}  that for the [HE, 3D] case, we obtain the best sensitivity
towards the NSI parameter $\varepsilon_{\mu\tau}$, whereas the [LE, 2D] mode gives the most conservative limits. 

 \begin{table}[htb!]
 \begin{center}
  \begin{tabular}{|l|l|l|l|}
  \hline\hline
   \multirow{2}{*}{Observable} & \multirow{2}{2em}{Binning scheme} & \multicolumn{2}{c|}{Constraints at 3$\sigma$ (90$\%$ C.L.)}  \\
   \cline{3-4}
                             &                                  &    \hspace{1.1cm}NH (true)    &  \hspace{1.1cm} IH (true)  \\
                             \hline
   \multirow{4}{*}{(E$_{\mu}$, $\cos\theta_{\mu}$)} & \multirow{2}{4em}{ LE} 
  &  $-0.06 <\varepsilon_{\mu\tau}< 0.07$   &  $-0.062 <\varepsilon_{\mu\tau}< 0.07$       \\
                                                    &                         
  &  ($-0.03 <\varepsilon_{\mu\tau}< 0.034$) & ( $-0.032<\varepsilon_{\mu\tau}<0.034$)          \\
                                                     \cline{2-4}
                                                    &\multirow{2}{4em}{ HE}  
  &  $-0.03 <\varepsilon_{\mu\tau}< 0.031$  & $-0.032<\varepsilon_{\mu\tau}< 0.032$         \\
                                                    &                         
  &  ($-0.016<\varepsilon_{\mu\tau}< 0.016$) & ($-0.016<\varepsilon_{\mu\tau}<0.016 $)         \\
\hline\hline
   \multirow{4}{*}{($E_\mu, \, \cos\theta_\mu,\,E'_{\rm had}$)} &  \multirow{2}{4em}{ LE}  
   &  $-0.028<\varepsilon_{\mu\tau}< 0.03$ & $0.03<\varepsilon_{\mu\tau}< 0.032$                        \\
                                                                  &                         
   &  ($-0.014<\varepsilon_{\mu\tau}< 0.014$) & ($-0.015 <\varepsilon_{\mu\tau}< 0.016$)  \\
                                                                  \cline{2-4}
                                                                  &   \multirow{2}{4em}{ HE} 
   &  $-0.018<\varepsilon_{\mu\tau}< 0.019$ & $-0.02 <\varepsilon_{\mu\tau}< 0.02 $                  \\
                                                                  &                          
   &  ($-0.01<\varepsilon_{\mu\tau}< 0.01$) & ($-0.01 <\varepsilon_{\mu\tau}< 0.01$)                      \\
  \hline\hline    
  \end{tabular}
  \end{center}
  \mycaption{The expected bound on $\varepsilon_{\mu\tau}$ for four different choices of 
  binning schemes and observables at $3\sigma$ and $90\%$ C.L. obtained using 500 kt$\cdot$yr exposure 
  of the ICAL detector. We give results for the both choices of true mass hierarchy. To obtain these constraints, 
  we marginalize over $\theta_{23}$, $\Delta m^{2}_{\rm eff}$, and both the choices of mass hierarchy in the fit. }
\label{tab:nsi-bound}
 \end{table}
The 3$\sigma$ (90$\%$) confidence level bounds on the flavor violating NSI parameter $\varepsilon_{\mu\tau}$ 
obtained using 500 kt$\cdot$yr exposure of the ICAL are listed in Table\,\ref{tab:nsi-bound}. The results are 
shown for true NH (3rd column) and true IH (4th column). For the [HE, 3D] case, we expect the best limit 
of $-0.01 <\varepsilon_{\mu\tau}< 0.01$ at 90$\%$ C.L. using 500 kt$\cdot$yr exposure of the ICAL detector and 
irrespective of the choices of true mass hierarchy. For the [LE, 2D] mode, we obtain the most conservative limit of 
$-0.03 < \varepsilon_{\mu\tau} < 0.034$ at 90$\%$ confidence level assuming NH as true choice. So far we have considered   
$\sin^2 2\theta_{13}= 0.1$ as our benchmark choice both in data and theory. If we consider the current best fit value of $\sin^2 2\theta_{13} = 0.085$\,\cite{Capozzi:2017ipn,Esteban:2018azc,deSalas:2017kay}, we have checked that our results will remain almost unaltered.  
For an instance, if we consider $\varepsilon_{\mu\tau} = 0.02$ in the fit, then we obtain $\Delta \chi^2 = 9.49$ assuming $\sin^2 2\theta_{13}= 0.1$ for the [HE, 3D] mode (see the red curve in the right panel of Fig.\,\ref{fig:bound-nh}). Under the same condition, if we take  $\sin^2 2\theta_{13} = 0.085$, then the $\Delta \chi^2$ changes to $9.38$.

\subsection{Advantage of having Charge Identification Capability}

 \begin{figure}[t!]
 \begin{center}
  \begin{minipage}{.49\linewidth}
      \includegraphics[width=\linewidth]{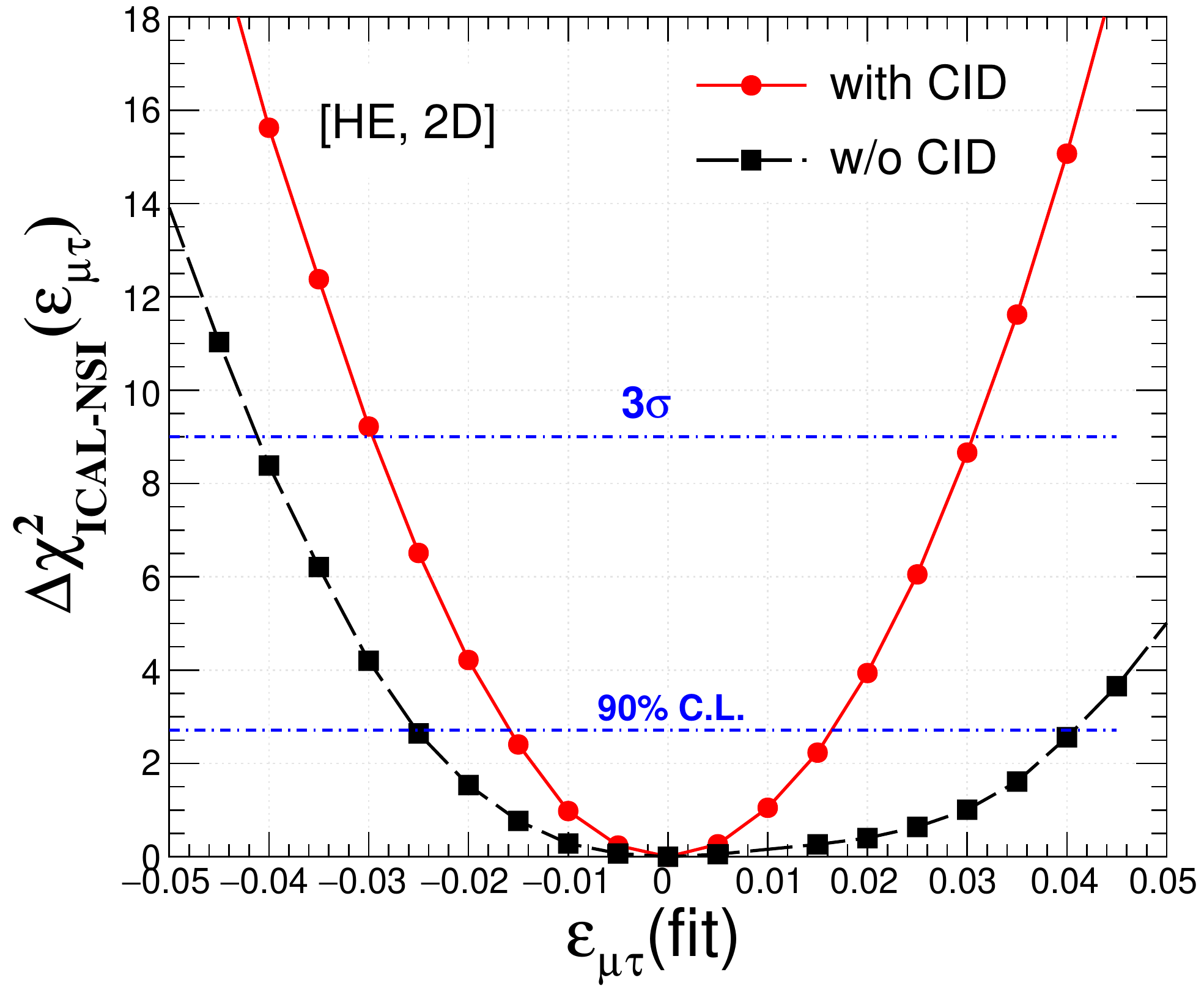}
  \end{minipage}
  \hfill
  \begin{minipage}{.49\linewidth}
      \includegraphics[width=\linewidth]{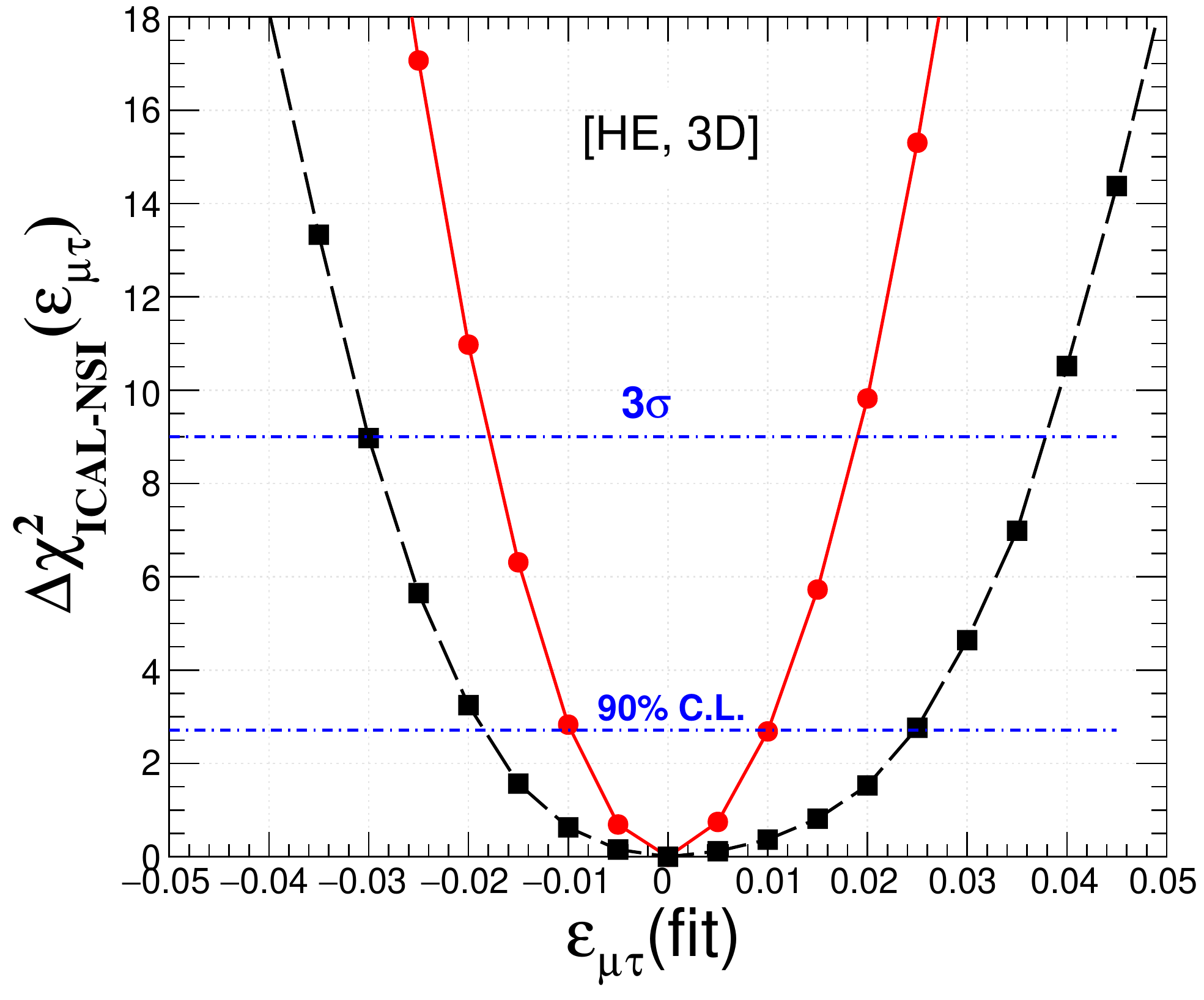}
  \end{minipage}  
  \mycaption{ In each panel, the red solid (black dashed) line shows the  expected sensitivity on $\varepsilon_{\mu\tau}$
  with (without) charge identification capability of ICAL. The left (right) panel is 
  for the 2D: $E_{\mu}$, $\cos \theta_{\mu}$ (3D: $E_{\mu}$, $\cos\theta_{\mu}$, $E'_{\rm had}$) mode 
  assuming the HE binning scheme. We consider 500 kt$\cdot$yr exposure and NH. Here, we keep all 
  the oscillation parameters fixed in the fit (fixed parameter scenario). } 
\label{fig:effect-cid-sen-he}
\end{center} 
\end{figure}
The ICAL detector is expected to have 
a uniform magnetic field of strength around 1.5 Tesla over the entire detector. 
It will enable the ICAL detector to identify the $\mu^-$ and $\mu^+$ events separately 
by observing the bending of their tracks in the opposite directions in the presence of 
the magnetic field. We label this feature of ICAL as the charge identification capability. 
In Ref.\,\cite{Chatterjee:2014vta}, it has been demonstrated that the ICAL detector 
will have a very good CID efficiency over a wide range of reconstructed $E_\mu$ and 
$\cos\theta_\mu$. In this work, we estimate for the first time the gain in the sensitivity 
that ICAL may have in constraining the NSI parameter $\varepsilon_{\mu\tau}$ due to its CID 
capability. In each panel of Fig.\,\ref{fig:effect-cid-sen-he}, we show the expected 
sensitivity of ICAL in constraining $\varepsilon_{\mu\tau}$ with (red solid line) and without 
(black dashed line) CID capability using 500 kt$\cdot$yr exposure and assuming NH. While 
preparing these plots, we keep the oscillation parameters fixed in the fit  and depict the 
result for the 2D: $E_{\mu}$, $\cos \theta_{\mu}$ (3D: $E_{\mu}$, $\cos\theta_{\mu}$, $E'_{\rm had}$) mode 
in the left (right) panel assuming the HE binning scheme.  It is apparent from Fig.\,\ref{fig:effect-cid-sen-he} 
that the CID capability of ICAL in distinguishing $\mu^-$ and $\mu^+$ events plays an important role to 
make it sensitive to the NSI parameter  $\varepsilon_{\mu\tau}$ like the mass hierarchy 
measurements\,\cite{Devi:2014yaa,Kumar:2017sdq}. In the following, we quote the 90$\%$ confidence level 
limits on $\varepsilon_{\mu\tau}$ that the ICAL detector can place with and without CID capabilities 
for [HE, 2D] and [HE, 3D] modes. 
\vskip 0.1 cm
\noindent
$\bullet$ [HE, 2D] mode (left panel of Fig.\,\ref{fig:effect-cid-sen-he}):
\begin{eqnarray}
{\rm with\,\,CID:}\,\,\hspace{.5cm}  & -0.015 
<\varepsilon_{\mu\tau}< 0.017  &\hspace{1cm}{\rm at}\,\, 
90\%\,{\rm C.L.} \,,\nonumber\\
{\rm without\,\,CID:}\,\,\hspace{.5cm}  & -0.025 
<\varepsilon_{\mu\tau}< 0.04    &\hspace{1cm}{\rm at}\,\, 
90\%\,{\rm C.L.} \,
\label{eq:bound-he-2d-cid}
\end{eqnarray}
$\bullet$ [HE, 3D] mode (right panel of 
Fig.\,\ref{fig:effect-cid-sen-he}):
\begin{eqnarray}
{\rm with\,\,CID:}\,\,\hspace{.5cm}  & -0.01 
<\varepsilon_{\mu\tau}< 0.011  &\hspace{1cm} {\rm at}\,\, 90\%\,{\rm C.L.} \,,\nonumber\\
{\rm without\,\,CID:}\,\,\hspace{.5cm} & -0.018 
<\varepsilon_{\mu\tau}< 0.025  &\hspace{1cm} {\rm at}\, \,90\%\,{\rm C.L.}\,
\label{eq:bound-he-3d-cid}
\end{eqnarray}
The limits on $\varepsilon_{\mu\tau}$ mentioned in Eq.\,\ref{eq:bound-he-2d-cid} and Eq.\,\ref{eq:bound-he-3d-cid} clearly demonstrate the improvement that the ICAL detector can have in constraining 
the NSI parameter $\varepsilon_{\mu\tau}$ due its CID capability.

\subsection{Limits on $\varepsilon_{\mu\tau}$ for various exposures}

\begin{center}
 \begin{figure}[t!]
   \begin{minipage}{.49\linewidth}  
   \includegraphics[width=\linewidth]{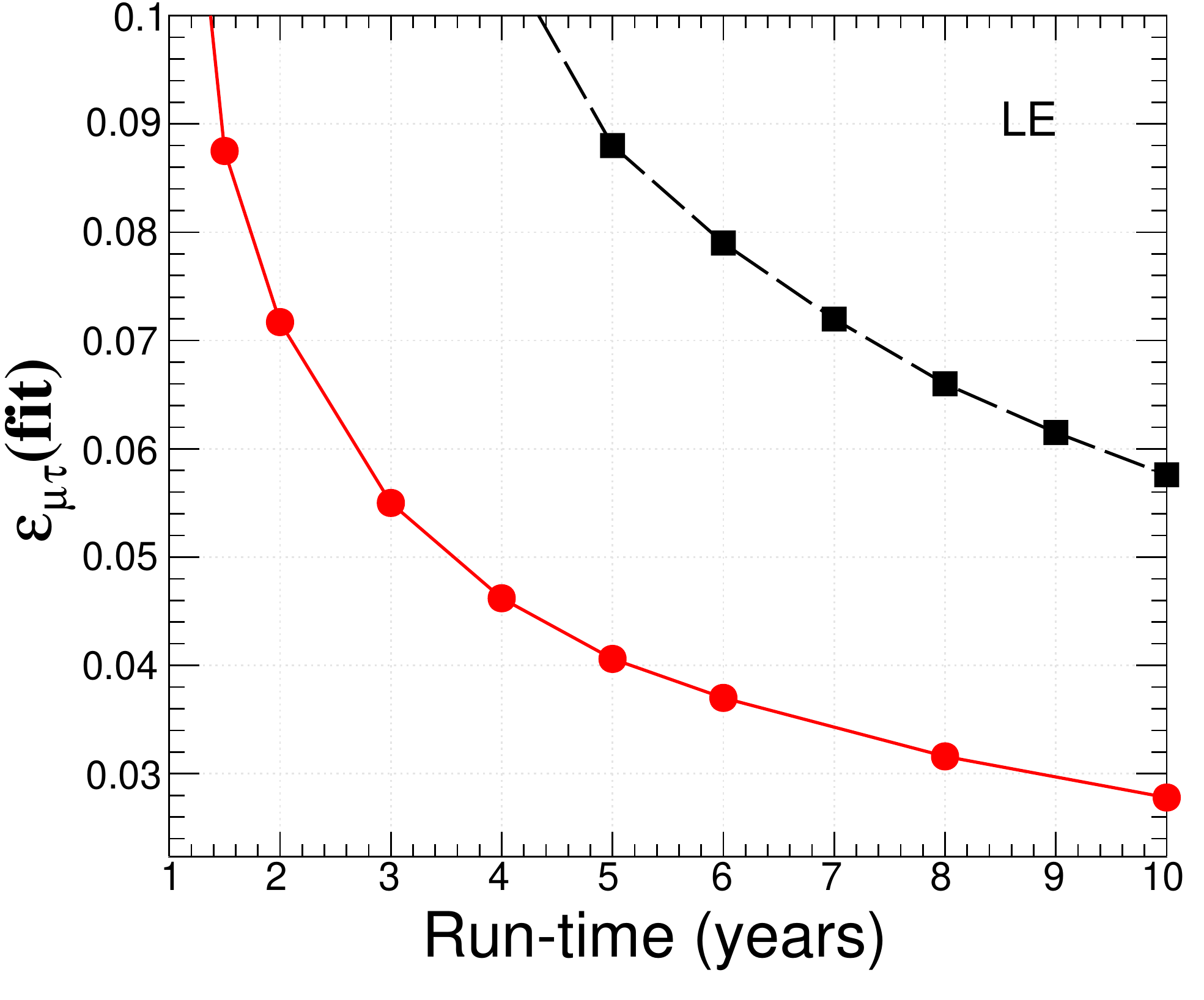}
 \end{minipage}
  \hfill
  \begin{minipage}{.49\linewidth}  
  \includegraphics[width=\linewidth]{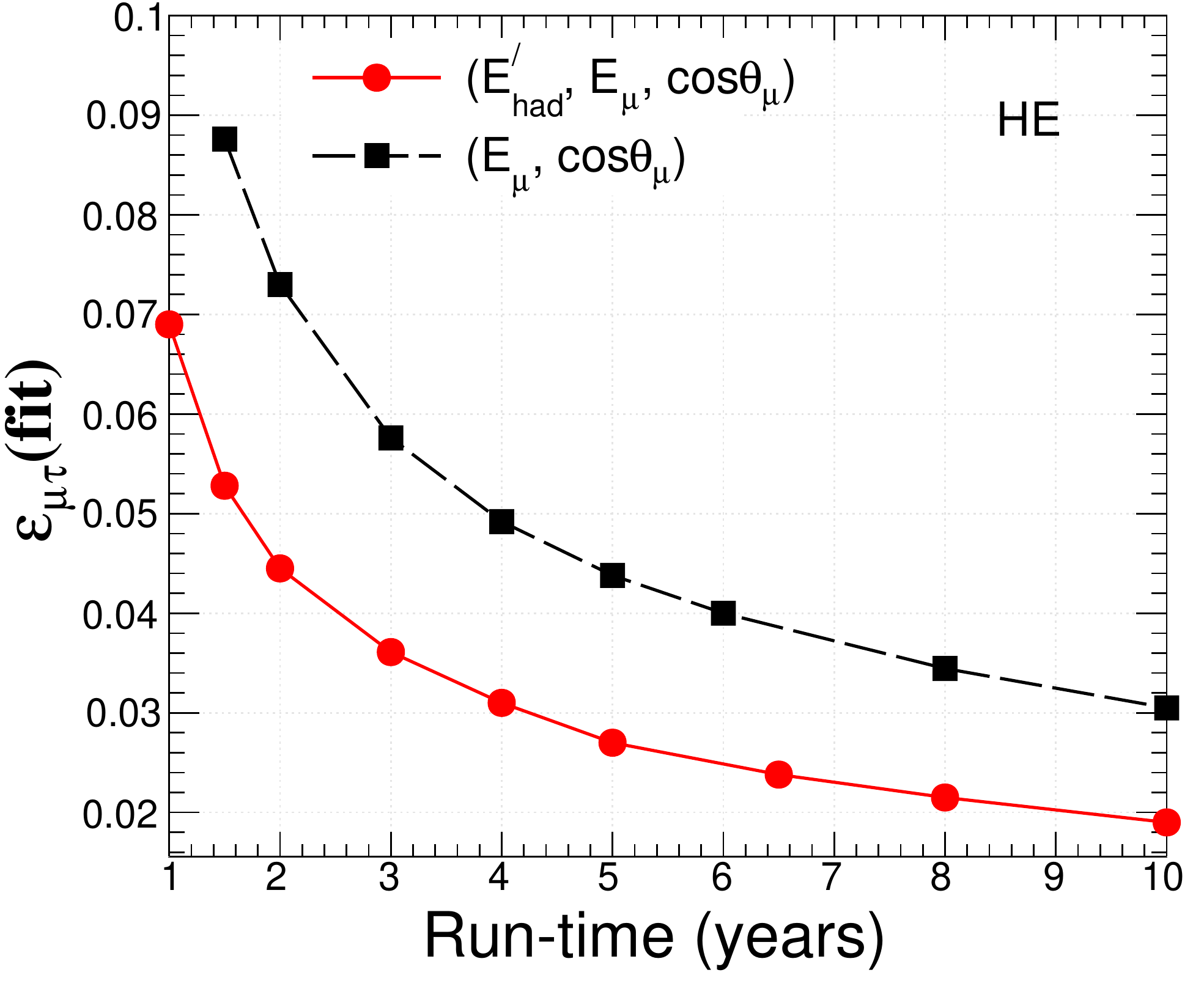}
  \end{minipage}
  \mycaption{Upper limits on $\varepsilon_{\mu\tau}$ at $3\sigma$ C.L. as a function of 
  run-time for 50 kt ICAL assuming NH and keeping all the oscillation parameters fixed in the fit. 
  Left (right) panel is for LE (HE) binning scheme. In each panel, the black and red lines depict 
  the results for 2D ($E_\mu$, $\cos\theta_\mu$) and 3D ($E_{\mu}$, $\cos\theta_{\mu}$, $E'_{\rm had}$) 
  modes respectively.
  }  
  \label{fig:runtime-epsilon}
  \end{figure}
\end{center}

Fig.\,\ref{fig:runtime-epsilon} shows the 3$\sigma$ limit on $\varepsilon_{\mu\tau}$ 
as a function of run-time\footnote{Note that while varying run-time in Fig.\,\ref{fig:runtime-epsilon}, 
we always consider the same LE and HE binning schemes as given in Table\,\ref{tab:bin-nsichap1} 
and Table\,\ref{tab:bin-nsichap2} respectively. For less exposure (small run-time), we may not have 
sufficient statistics in most of the bins. One needs to consider larger bin widths to tackle this issue 
which in turn may effect the sensitivity results. We have plans to address this issue in our future study.} 
for  50 kt ICAL. The left (right) panel is for LE (HE) binning scheme. In each panel,  the black and red 
lines depict the results for 2D ($E_\mu$, $\cos\theta_\mu$) and 3D ($E_{\mu}$, $\cos\theta_{\mu}$, $E'_{\rm had}$) 
modes respectively. In Fig.\,\ref{fig:runtime-epsilon}, various sensitivity curves are drawn keeping 
all the oscillation parameters fixed in the fit and assuming NH. Here, we give the results only for 
positive values of $\varepsilon_{\mu\tau}$. We have checked that the results look similar if we consider 
negative values of $\varepsilon_{\mu\tau}$ as well. If we take total 250 kt$\cdot$yr exposure 
(50 kt ICAL with a run-time of 5 years), the expected bound 
is $|\varepsilon_{\mu\tau}|\lesssim 0.28$ at $3\sigma$ C.L. assuming NH in [HE, 3D] binning scheme. 
It suggests that ICAL can place competitive constraints on $\varepsilon_{\mu\tau}$ even for less exposure.

\subsection{Impact of non-zero $\varepsilon_{\mu\tau}$ on Mass Hierarchy Determination} 
\label{sec:mh-label}

This section is devoted to study how  the flavor violating NSI parameter $\varepsilon_{\mu\tau}$ 
affects the mass hierarchy measurement which is the prime goal of the ICAL detector. 
We quantify the performance ICAL to rule out the wrong hierarchy by adopting  
the following $\chi^2$ expression: 
\begin{equation}
 \Delta \chi^2_{\rm ICAL-MH} = \chi^2_{\rm ICAL} 
 ({\rm false\,\, MH}) - \chi^2_{\rm ICAL} ({\rm true \,\, MH})\,.
 \label{eq:ical-mh-sen}
\end{equation}
Here, we obtain  $\chi^2_{\rm ICAL} ({\rm true\,\,MH})$  
and $\chi^2_{\rm ICAL} ({\rm false\,\,MH})$ by performing the 
fit to the prospective data assuming true and false mass 
hierarchy respectively. Since the statistical fluctuations 
are suppressed in our analysis, $\chi^2_{\rm ICAL} 
({\rm true\,MH}) \approx 0$. First, we estimate the sensitivity 
of the ICAL detector to determine the neutrino mass hierarchy
by adopting the procedure as outlined in Ref.\,\cite{Devi:2014yaa} 
for the standard case, which we denote as 
``$\Delta\chi^2_{\rm{ICAL-MH}}$\,(SM)'' in the third column of 
Table\,\ref{tab:MO-sen}. Next, to estimate the mass 
hierarchy sensitivity in the presence of non-zero 
$\varepsilon_{\mu\tau}$, we adopt the following strategy.  
\begin{table}[t!]
\begin{center}
 \begin{tabular}{|c|c|c|c|c|}
 \hline
True MH & \hspace{2mm}Analysis Mode\hspace{2mm}  & $\Delta\chi^2_{\rm{ICAL-MH}}$\,(SM) 
&  $\Delta\chi^2_{\rm{ICAL-MH}}$\,(SM + $\varepsilon_{\mu\tau}$) & Reduction  \\
\hline
\hline
\multicolumn{5}{|c|}{LE binning scheme} \\
\cline{1-5}
NH & \makecell[c]{ ($E_\mu , \, \cos\theta_\mu$) \\($E_\mu, \, \cos\theta_\mu, \, E'_{\rm had}$) }
    & \makecell[c] {5.62 \\8.66}  & \makecell{ 4.81\\7.49} & \makecell[c]{14.4$\%$ \\ 13.5$\%$ }\\
\hline
IH & \makecell[c]{ ($E_\mu , \, \cos\theta_\mu$) \\($E_\mu, \, \cos\theta_\mu, \, E'_{\rm had}$) }
    &\makecell[c]{ 5.31\\8.48 } &  \makecell{4.14 \\ 6.88 } &\makecell[c]{22.0$\%$ \\ 18.9$\%$ } \\ 
\hline
\hline
\multicolumn{5}{|c|}{HE binning scheme}\\
\hline
NH & \makecell[c]{ ($E_\mu , \, \cos\theta_\mu$) \\($E_\mu, \, \cos\theta_\mu, \, E'_{\rm had}$) }
    & \makecell[c] {5.96 \\9.13}  & \makecell{ 5.37\\ 8.16} & \makecell[c]{9.9$\%$ \\ 10.6$\%$ }\\
\hline
IH & \makecell[c]{ ($E_\mu , \, \cos\theta_\mu$) \\($E_\mu, \, \cos\theta_\mu, \, E'_{\rm had}$) }
    &\makecell[c]{ 5.66\\ 8.99 } &  \makecell{4.95 \\ 7.66 } &\makecell[c]{12.5$\%$ \\ 14.8$\%$ } \\
 \hline
 \end{tabular}
\end{center}
\mycaption{The mass hierarchy sensitivity of the ICAL detector 
using 500 kt$\cdot$yr exposure.  For the ``SM'' case (third 
column), we do not consider $\varepsilon_{\mu\tau}$ in data 
and in fit. For the ``SM +  $\varepsilon_{\mu\tau}$'' case 
(fourth column), we introduce $\varepsilon_{\mu\tau}$ in the 
fit and marginalize over it in the range of [-0.1, 0.1] along 
with oscillation parameters $\theta_{23}$ and $\Delta m^2_{\rm eff }$. 
Last column shows how much the mass hierarchy sensitivity deteriorates 
in presence of $\varepsilon_{\mu\tau}$ as compared to the SM case. 
We present our results for various choices of binning schemes and observables 
assuming both true NH and true IH.  }
\label{tab:MO-sen}
\end{table}
We generate the data with a given mass hierarchy assuming 
$\varepsilon_{\mu\tau} = 0$. Then,  while fitting the 
prospective data with the opposite hierarchy, we introduce 
$\varepsilon_{\mu\tau}$ in the fit and marginalize over it 
in the range of - 0.1 to 0.1  along with the oscillation 
parameters  $\theta_{23}$ and $\Delta m^2_{\rm eff }$ in their 
allowed ranges as mentioned in 
Sec.\,\ref{sec:method-analysis-nsi}. We label this result as 
``$\Delta\chi^2_{\rm{ICAL-MH}}$\,(SM + $\varepsilon_{\mu\tau}$)'' 
in the fourth column of Table\,\ref{tab:MO-sen}. 
We show our results for various choices of binning schemes and observables
assuming both true NH and true IH. We consider 500 kt$\cdot$yr exposure of 
the ICAL detector. We can see from Table\,\ref{tab:MO-sen} that depending 
on the choice of true mass hierarchy and the analysis mode, the mass hierarchy 
sensitivity of ICAL gets reduced by 10$\%$ to 20$\%$ due to the presence of non-zero 
$\varepsilon_{\mu\tau}$ in the fit. 

\subsection{Precision Measurement of Atmospheric Parameters with non-zero $\varepsilon_{\mu\tau}$}

\begin{figure}[t!]
 \begin{center}
  \includegraphics[width=.8\textwidth]{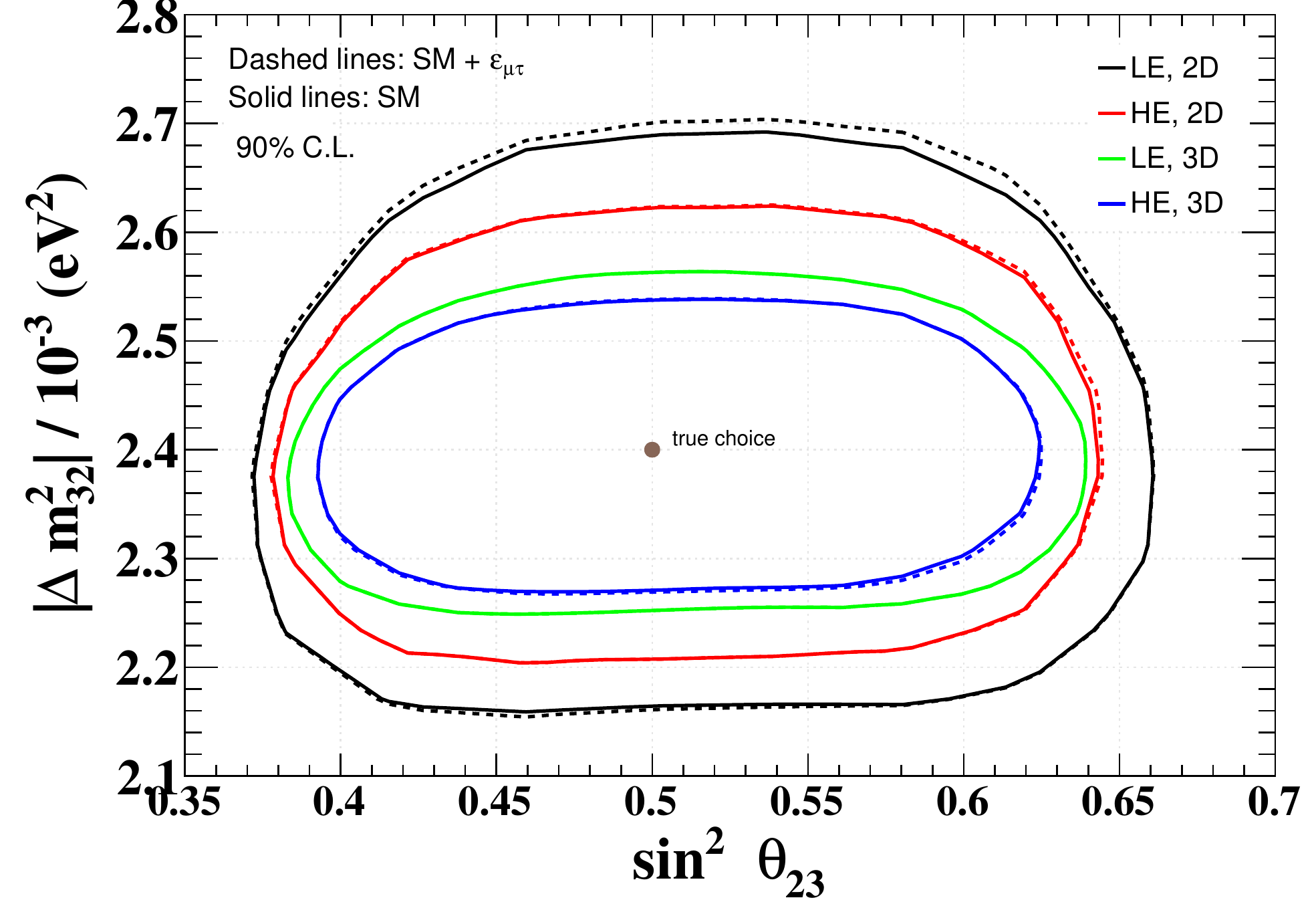}
  \mycaption{90$\%$ C.L. (2 d.o.f.) allowed regions in 
  $\sin^2\theta_{23}$ - $|\Delta m^2_{32}|$ plane for 500 kt$\cdot$yr 
  exposure of the ICAL detector assuming NH. The brown dot represents the 
  true choices of $\sin^2\theta_{23}$ and $|\Delta m^2_{32}|$.  
  The solid lines show the results for the ``SM'' case, where we 
  do not consider $\varepsilon_{\mu\tau}$ in data and in fit. The dashed 
  lines portray the results when we introduce $\varepsilon_{\mu\tau}$ in 
  the fit and marginalize  over its  $\pm$10$\%$ range. For other details, 
  see text. }
  \label{fig:nsi-pm}
 \end{center} 
\end{figure}
Next, we turn our attention to the precise measurement of 
atmospheric oscillation parameters $\sin^2\theta_{23}$ and 
$|\Delta m^2_{32}|$  using 500 kt$\cdot$yr exposure of 
the ICAL detector.  We quantify this performance indicator 
using the following expression:
\begin{equation}
  \Delta \chi^2_{\rm ICAL-PM}\, \big(\sin^2\theta_{23},\,
  |\Delta m^2_{32}|\big) \,\,= \,\, \chi^2_{\rm ICAL} \big(\sin^2\theta_{23},\,
  |\Delta m^2_{32}|\big)  \,\, - \,\, \chi^2_0 \,, 
 \label{eq:ical-pm-sen-nsi}
\end{equation}
where   $\chi^2_0$ is the minimum value of $\chi^2_{\rm ICAL}$ in 
the allowed  parameter range. Since we suppress the statistical fluctuations, we have 
$\chi^2_0 \approx 0$. First, considering $\sin^2\theta_{23}$\,(true)\,=\,0.5 
and $|\Delta m^2_{32}|$ (true) = 2.4 $\times$ 10$^{-3}$ eV$^2$, 
we estimate the allowed regions in $\sin^2\theta_{23}$ -
$|\Delta m^2_{32}|$ (test) plane in the absence of 
$\varepsilon_{\mu\tau}$ at 90$\%$ C.L. (2 d.o.f.). 
We show these results for the ``SM'' case using solid lines in 
Fig.\,\ref{fig:nsi-pm} for various analysis modes. For the [HE, 3D] case, 
we achieve the best precision for the atmospheric parameters, and for the 
[LE, 2D] case, we have the most conservative results. 

Next, we study the impact of non-zero $\varepsilon_{\mu\tau}$
in the precision measurement of atmospheric parameters in the 
following fashion. We again generate the prospective data 
considering the true values of $\sin^2\theta_{23}$ and $|\Delta m^2_{32}|$ 
as mentioned above. Then, while estimating the allowed regions in 
 $\sin^2\theta_{23}$ - $|\Delta m^2_{32}|$ (test) plane, we 
 introduce $\varepsilon_{\mu\tau}$ in the fit and marginalize 
 over it in the range of [-0.1,\,0.1]. We present these results 
 for the ``SM + $\varepsilon_{\mu\tau}$'' case
at $90\%$ C.L. (2 d.o.f.) with the help of dashed lines in  
Fig.\,\ref{fig:nsi-pm} for various analysis modes. We do not 
see any appreciable change in the contours when we introduce 
$\varepsilon_{\mu\tau}$ in the fit and vary in its $\pm$10$\%$ 
range. It suggests that the precision measurement of atmospheric
oscillation parameters at the ICAL detector is quite robust 
even if we marginalize over $\varepsilon_{\mu\tau}$ in the fit.
Similar results were obtained by the Super-Kamiokande Collaboration 
in Ref.\,\cite{Mitsuka:2011ty}, where they studied the impact of NSI's 
in $\nu_\mu$-$\nu_\tau$ sector using their Phase I and Phase II 
atmospheric data.

\section{Summary and Concluding Remarks}
\label{sec:conclusions-nsichap}
In this paper, we explore the possibility of lepton flavor violating neutral current 
non-standard interactions (NSI's) of atmospheric neutrino and antineutrino 
while they travel long distances inside the Earth matter before reaching to the ICAL detector.
During the propagation of these neutrinos, we allow an extra interaction vertex with $\nu_\mu$ as 
the incoming particle and $\nu_\tau$ as the outgoing one and vice versa.
With such an interaction vertex, the neutral current non-standard interaction of neutrino with matter 
fermions gives rise to a new matter potential whose relative strength as compared to 
the standard matter potential ($V_{CC}$) is denoted by $\varepsilon_{\mu\tau}$. 

We show 
that the ICAL detector would be able to place tight constraints on the NSI parameter 
$\varepsilon_{\mu\tau}$ considering reconstructed hadron energy and muon momentum as 
observables. We find that with $E_\mu \in [1,11]$\,GeV and with [$E_\mu,\,\cos\theta_\mu$] 
as observables, the expected limit on $\varepsilon_{\mu\tau}$ at 90$\%$ C.L. is 
$-0.03 < \varepsilon_{\mu\tau} <0.03$. If we increase the muon energy range from 11 to 
21 GeV ($E_\mu \in [1, 21]$\,GeV) and consider the reconstructed hadron energy 
($E'_{\rm had}$) as an extra observable on top of the four momenta of muon 
($E_\mu,\,\cos\theta_\mu$), we find a significant improvement in the limit 
which is $-0.01 <  \varepsilon_{\mu\tau} < 0.01 $ at 90$\%$ C.L. using 500 
kt$\cdot$yr exposure of the ICAL detector. We observe that the charge identification 
capability of the ICAL detector plays an important role to obtain these
tight constraints on  $\varepsilon_{\mu\tau}$  as mentioned above. 

Assuming 1 to 21 GeV reconstructed muon energy range and considering 
$E_\mu,\,\cos\theta_\mu,\,$and $E'_{\rm had}$ as 
observables, we find that the mass hierarchy sensitivity at the ICAL 
detector deteriorates by $\sim$10$\%$ 
if we introduce the NSI parameter $\varepsilon_{\mu\tau}$ in the fit 
and marginalize over it in the range of 
-0.1 to 0.1 along with other standard oscillation parameters.   
On the other hand, the precision measurement of atmospheric 
oscillation parameters at the ICAL detector is  
quite robust even if we marginalize over the NSI parameter 
$\varepsilon_{\mu\tau}$ in fit in the range -0.1 to 0.1.

\section{Acknowledgment}

The INO-ICAL Collaboration is exploring various physics 
potentials of the proposed ICAL detector and this work
is a part of that ongoing effort. This work would not have 
been possible without the contribution of several members 
of the Collaboration. We would like to thank A. Dighe, A.M. 
Srivastava, P. Agrawal, S. Goswami, D. Indumathi, S. Choubey, 
S. Uma Sankar for their useful comments on our work.
We thank N. Mondal, A. Dighe, and S. Goswami for helping 
us during the INO Internal Review Process. A.K. acknowledges 
the support from the Department of Atomic Energy (DAE), 
Government of India. S.K.A. acknowledges the support from 
DST/INSPIRE Research Grant [IFA-PH-12], Department of Science 
and Technology, India and the Young Scientist Project 
[INSA/SP/YSP/144/2017/1578] from the Indian National 
Science Academy. T. T. acknowledges support from the 
Ministerio de Econom\'{\i}ay Competitividad (MINECO): 
Plan Estatal de Investigaci\'{o}n (ref. FPA2015- 65150-C3-1-P, 
MINECO/FEDER), Severo Ochoa Centre of Excellence 
and MultiDark Consolider (MINECO), and Prometeo program 
(Generalitat Valenciana), Spain.


\bibliographystyle{JHEP} 
\bibliography{NSI-Atmospheric.bib}

\end{document}